\documentclass[12pt]{article}
\usepackage{amsmath,amssymb,exscale,multirow,graphicx}
\usepackage[utf8]{inputenc}
\usepackage{xcolor}
\usepackage{cite}
\usepackage{hyperref}

\input epsf
\newcommand{\m}[1]{\marginpar{{\tiny *}} }
\newcommand{\pslash}{{\not \!p}}
\def\rd{R_{D^{(*)}}}

\def\det{{\rm det}}

\def\be{\begin{equation}}
\def\te{\end{equation}} 
\def\nn{\nonumber}

\def\0{{(0)}}

\def\al#1\fal{\begin{align}#1\end{align}} 
 
\def\d{\delta}

\def\d1{\delta}

\newcommand\xs[1]{x_{L,#1}}

\begin{document}
\topmargin -1.0cm
\oddsidemargin -0.8cm
\evensidemargin -0.8cm

\begin{center}
\vspace{40pt}

\Large \textbf{The flavor of a minimal composite $S_1$ leptoquark \\ and the $b\to c\tau\bar\nu$ anomaly}

\end{center}

\vspace{15pt}
\begin{center}
{\bf Leandro Da Rold$^{\star}$} 

\vspace{20pt}

\textit{Centro At\'omico Bariloche, Instituto Balseiro and CONICET}
\\[0.2cm]
\textit{Av.\ Bustillo 9500, 8400, S.\ C.\ de Bariloche, Argentina}

\end{center}

\vspace{20pt}
\begin{center}
\textbf{Abstract}
\end{center}
\vspace{5pt} {\small \noindent
Several experiments have measured a deviation in $B\to D^{(*)}$ semileptonic decays, that point to new physics at the TeV scale violating lepton flavor universality. A scalar leptoquark $S_1$, with a suitable structure of couplings in flavor space, is known to be able to solve this anomaly modifying $b\to c\tau\bar\nu$. In the context of composite Higgs models, we consider a theory containing $H$ and $S_1$ as Nambu-Goldstone bosons (NGBs) of a new strongly interacting sector, with ordinary resonances at a scale ${\cal O}(10)$~TeV. Assuming anarchic partial compositeness of the Standard Model (SM) fermions we calculate the potential of the NGBs that is dominated by the fermions of the third generation, we compute $\rd$ and estimate the corrections to flavor observables by the presence of $S_1$. We find that the SM spectrum and $m_{S_1}\sim$~TeV can be obtained with a NGB decay constant of order $\sim 5$~TeV.
We obtain a robust correlation between the main corrections to $\rd$, $B_{K^{(*)}\nu\nu}$ and $g_\tau/g_\mu$, that leads to a sever bound on $\rd$, roughly $2\sigma$ below the experimental value. Besides the bounds on the flavor observables $g_\tau^W$, BR$(\tau\to\mu\gamma)$ and $\Delta m_{B_s}$ are saturated, with the first one requiring a coupling between resonances $g_*\lesssim 2$, whereas the second one demands $m_{S_1}\gtrsim 1.7$~TeV, up to corrections of ${\cal O}(1)$.
}

\vfill
\noindent {\footnotesize E-mail:
$\star$ daroldl@ib.edu.ar
}

\noindent
\eject

\tableofcontents
\section{Introduction}
The LHC is confirming the enormous success of the Standard Model of particle physics, with no direct evidence of new particles at the few TeV scale up to know. However in the last years several experiments have measured deviations with respect to the SM in the decay of the $B$-meson to $D\tau\bar\nu$ and $D^*\tau\bar\nu$, compared with the decays to light leptons $\ell=e,\mu$, parameterized in the ratio of branching ratios $\rd$~\cite{babar1,BaBar:2012obs,RD_Babar,Belle:2015qfa,RD_Belle,Belle:2017ilt,RD_Belle2,Belle:2019rba,RD_LHCB,LHCb:2017smo,RD_LHCB2}. These deviations point towards a violation of lepton flavor universality (LFU) in the third generation, specifically in the decay $b\to c\tau\bar\nu$~\cite{London:2021lfn}.~\footnote{There have also been deviations in $B$-meson decays with neutral currents, but since the last measurements from LHCb agree with the SM~\cite{LHCb:2022qnv,LHCb:2022vje}, we will not consider deviations in those decays in this article.}

Since a confirmation of the violation of LFU in $\rd$ would be a signal of new physics, there has been a lot of activity in the last years, from the perspective of effective theories adding higher dimensional operators to the SM, as well as with the introduction of new states and interactions.~\footnote{Since a comprehensive list of articles on this topic is beyond the scope of our work, we refer to Refs.~\cite{London:2021lfn} and~\cite{EFT_Bhattacharya,EFT_Azatov,Alonso:2015sja,EFT_Greljo,EFT_Calibbi,EFT_Bordone,Huang:2018nnq} and references therein, and along the text we will add more references directly related with our article.} One of the best scenarios for the explanation of the deviation on $\rd$ is the presence of leptoquarks (LQs), in this work we focus on $S_1$, a spin-0 state with electric charge -1/3 and singlet of SU(2)$_L$~\cite{Sakaki:2013bfa,Alonso:2015sja,Freytsis:2015qca,Bauer:2015knc,Li:2016vvp,Hiller:2016kry,Popov:2016fzr,Cai:2017wry,Isidori,Angelescu:2018tyl,Iguro:2018vqb,Blanke:2018yud,Blanke:2019qrx,NP_Crivellin2,Bigaran:2019bqv,Crivellin,Cheung:2020sbq,Bordone:2020lnb,Lee:2021jdr,Angelescu:2021lln,Marzocca:2021miv,Belanger:2021smw,Sahoo:2021vug,Bhaskar:2022vgk,Li:2022chc,Chen:2022hle,Lizana:2023kei}. An $S_1$ at the TeV scale with 
couplings wit a particular structure of flavor, has been shown to provide one of the best fits to the experimental observations~\cite{2008.09548,Iguro:2022yzr}.
In this work we propose a model where the mass of $S_1$ can be naturally lighter than the scale of new physics, that has a rationale determining the flavor structure of its couplings,
and we study the predictions for flavor physics, as well as the possibility of an explanation of $\rd$.

Since new physics at the TeV scale is also required to stabilize the Higgs potential, we will consider that $S_1$ and the Higgs are the lightest states of a new strongly coupled sector (SCFT) that produces resonances at few TeV energies. To avoid conflict with precise electroweak (EW) measurements we will assume that the new resonances have masses of ${\cal O}(10)$~TeV, with $H$ and $S_1$ being Nambu-Goldstone bosons of the SCFT~\cite{Gripaios:2014tna,Alvarez:2018gxs,DaRold:2018moy,DaRold:2020bib}. A potential for the NGBs will be generated at radiative level by the interactions with the elementary fields of the SM, that explicitly break the global symmetry of the SCFT. We will compute the potential at one loop level and study the predictions for EW symmetry breaking, the masses of $H$ and $S_1$, and the mass of the top-quark.

Generically one can expect that $S_1$ interacts also with the rest of SM fermions, potentially introducing dangerous corrections to flavor observables. In this work we will assume the paradigm of anarchic partial compositeness, that has been shown to be compatible with most flavor observables for resonances at a scale ${\cal O}(10)$~TeV~\cite{Bauer:2009cf,Panico:2015jxa}. We will show that the model naturally selects Left-handed couplings to dominate the effect on $\rd$,
$B_{K^{(*)}\nu\nu}$ and $g_\tau/g_\mu$, and that it predicts a correlation between the corrections to these obsrvables, such that the $B$-anomalies can not be explained.
We will also compute the corrections from $S_1$ to low energy flavor observables, finding that a few of them are saturated. 

We will consider an effective description of the SCFT at energies below the scale of the massive resonances, based on the symmetries of the model. In order to be able to make more precise calculations, both of the spectrum of states as well as of some flavor observables, we will also consider an effective description of the first level of resonances of the SCFT by a 2-site model. For ultraviolet complete theories containing light LQs we refer the reader to Refs.~\cite{NP_Marzocca,NP_Becirevic2}. 

The article is organized as follows: in sec.~\ref{sec-model} we describe the model, its symmetries and the pattern of symmetry breaking, its flavor structure, the effective theory below the scale of resonances, the potential and a 2-site theory describing the elementary fields and the resonances. In sec.~\ref{sec-flavorobs} we compute the corrections to $\rd$ and other flavor observables, in sec.~\ref{sec-num} we show the numerical results, including the values of $\rd$ and $m_{S_1}$, in sec.~\ref{sec-resonances} we describe the resonances of the model and we conclude in sec.~\ref{sec-conclusions}.

\section{A model with $S_1$ and $H$ as composite NGBs}
\label{sec-model}
In this section we build a model that contains a composite $S_1$ LQ with a mass of order TeV, a composite Higgs instead of the elementary one of the SM, and a number of composite resonances with masses ${\cal O}(10)$~TeV. The fermions and gauge bosons of the SM are assumed to be elementary fields. 
The $S_1$ interactions with the elementary fermions are:
\begin{equation}
{\cal L}_ {\rm int} = x_L S_1 \bar q^c_L \tilde \epsilon l_L + x_R S_1 \bar u^c_R e_R \ .
\label{eq-iS1}
\end{equation}
where $\tilde \epsilon=i\sigma_2$, the real anti-symmetric tensor in two dimensions, $x_L$ and $x_R$ are dimensionless couplings and generation indices are understood for the fermions and couplings.
Generically there could also be terms with di-quarks that violate baryon number, however they will be forbidden by the symmetries of the model.

There are two important issues that we want to address in the article: in the first place having a separation between the masses of the composite resonances of the SCFT, as is the sought case for $H$ and $S_1$ and the rest of the states, is in general problematic, since generically all the masses receive quantum corrections of the same order. The lightness of $H$ and $S_1$ compared with the rest of the resonances will be ensured by making them NGBs of the SCFT, while the rest of the resonances will be ordinary states. Thus we assume that the SCFT has a global symmetry, spontaneously broken by its dynamics to a small group and generating only $H$ and $S_1$ as NGBs, realizing the minimal model. A potential for the NGBs will be generated by the interactions with the elementary sector that explicitly break the global symmetry of the SCFT. The separation between $m_{H}$ and $m_{S_1}$ will rely on the Higgs vacuum expectation value (vev) being smaller than the decay constant of the NGBs: $v_H/f_\Pi\sim{\cal O}(10^{-1})$. As is well known this separation of scales requires some tuning, that has been estimated to be of order $(v_H/f_\Pi)^2$~\cite{Agashe-2004,1210.7114}. 

On the second place a solution of $\rd$ requires sizable coupling of $S_1$ with the third generation of SM fermions, while a large amount of other low energy flavor observables give very stringent bounds on couplings generating flavor transitions or violating flavor universality. Instead of assuming couplings $x_L$ and $x_R$ with an ad-hoc structure in flavor space, we will consider the paradigm of anarchic partial compositeness (APC), that naturally leads to a pattern of flavor couplings related with the fermion masses~\cite{Gherghetta:2000qt,Agashe:2004cp}.

Partial compositeness of the SM fermions will be obtained with linear interactions between the elementary fermions and the operators of the SCFT: ${\cal L}_ {\rm int} = \omega_f \bar f {\cal O}_f$, with $\omega_f$ the coupling at a high energy scale $\Lambda$, $f$ any of the elementary fermions and ${\cal O}_f$ the SFCT operator. Assuming that at a scale $m_*\sim{\cal O}(10)$~TeV ${\cal O}_f$ generates massive resonances, a linear mixing is obtained:
\begin{equation}
{\cal L}_ {\rm pc} = \lambda_f f_\Pi \bar f F + {\rm h.c.} \ ,
\label{eq-pc}
\end{equation}
where we used a capital letter for the effective field of the resonance, $F$. $\lambda_f$ is the linear coupling at scale $m_*$, that is obtained by evolving $\omega_f$ with the RGEs. If $\Lambda\gg m_*$ and the SCFT has an approximate scale invariance, the running of the coupling is driven by the anomalous dimension of ${\cal O}_f$, such that for dim$({\cal O}_f)>5/2$ an exponentially small $\lambda_f$ can be generated~\cite{Kaplan:1991dc,Contino:2004vy}.

The resonances are characterized by their masses: $m_*$, and their couplings: $g_*\sim m_*/f_\Pi$, that are taken to be all of the same order: $1< g_*\ll 4\pi$, .

\subsection{Symmetries of the model}
\label{sec-sym}
Let us now consider in some detail the symmetries of the theory, such that the picture described in the previous section can emerge. Eq.~(\ref{eq-pc}) requires that the SCFT has a global symmetry containing the SM gauge symmetry, G$_{\rm SM}$. Besides we extend it with an extra global SU(2) factor, such that the composite sector has custodial symmetry. We call H$_{\rm min}$ to the minimal group satisfying those conditions: H$_{\rm min}={\rm SU}(3)_c\times{\rm SU}(2)_L\times{\rm SU}(2)_R\times{\rm U}(1)_X$, the factor U(1)$_X$ is required to properly normalize hypercharge of the SM fermions.

The SCFT is assumed to have a global symmetry group SO(11) spontaneously broken to SO(10), with a coset containing $S_1$ and $H$ as NGBs, as well as the unbroken subgroup containing H$_{\rm min}$~\footnote{This coset has also been considered in Refs.\cite{Frigerio:2011zg,Hosotani:2015hoa}, although in a different context, see also Refs.~\cite{Aydemir:2018cbb,Becirevic:2018afm,Aydemir:2019ynb} for scalar LQs in GUT.}. The coset of the NGBs is: 
\begin{equation}
{\rm SO}(11)/{\rm SO}(10)~\sim {\bf 10} \ .
\end{equation}
The spontaneous breaking, and the inclusion of H$_{\rm min}$ into SO(10), can be summarized as:
\begin{equation}
{\rm SO}(11)\to {\rm SO}(10) \supset {\rm SO}(6)\times {\rm SO}(4) \supset {\rm SU}(3)_c\times{\rm U}(1)_X\times{\rm SO}(4) \ ,
\label{eq-subg}
\end{equation}
with the algebra of ${\rm SO}(4)$ equivalent to the algebra of ${\rm SU}(2)_L\times{\rm SU}(2)_R$ and the inclusion ${\rm SU}(3)_c\times{\rm U}(1)_X\subset {\rm SO}(6)$ being straightforward since ${\rm SO}(6)\simeq {\rm SU}(4)$.

Under the sequence of subgroups of Eq.~(\ref{eq-subg}), the NGBs decompose as:
\begin{equation}
{\bf 10}\sim ({\bf 6},{\bf 1},{\bf 1})\oplus({\bf 1},{\bf 2},{\bf 2})\sim ({\bf 3},{\bf 1},{\bf 1})_2\oplus(\bar{\bf 3},{\bf 1},{\bf 1})_{-2}\oplus({\bf 1},{\bf 2},{\bf 2})_0 \ .
\label{eq-NGBdec}
\end{equation}
Identifying hypercharge as:
\begin{equation}
Y=T^{3R}-X/6 \ ,
\label{eq-Y}
\end{equation}
the multiplets of the r.h.s of Eq.~(\ref{eq-NGBdec}) have the right transformation properties under G$_{\rm SM}$ to be identified with $S_1$ and its conjugate and $H$. The coset is minimal in the sense that there are no extra NGBs, and the ranks of the groups involved in the coset are the smallest ones containing H$_{\rm min}$.

The vacuum and the NGBs can be parameterized as:
\begin{align}
& \phi = \Sigma\phi_0 \ , \qquad \Sigma=e^{i\Pi/f_\Pi} \ , \qquad \Pi=\Pi^{\hat a}T^{\hat a} \ , 
\nonumber
\\
& \phi_0=(0,0,0,0,1,0,0,0,0,0,0) \ ,
\label{eq-U}
\end{align}
with $T^{\hat a}$ the broken generators, $\Pi^{\hat a}$ the NGB fields and $f_\Pi$ their decay constant. A vev of the NGBs produces a vacuum misalignment that, by using the generators of Ap.~\ref{ap-so11}, can be conveniently parameterized as:
\begin{align}
&\langle\phi\rangle = (0,0,0,\epsilon,\sqrt{1-\epsilon^2-\theta^2},\theta,0,0,0,0,0) \ , 
\label{eq-vevg1}
\\
&\epsilon=\frac{v_H}{\sqrt{v_H^2+v_{S_1}^2}}\sin\frac{\sqrt{v_H^2+v_{S_1}^2}}{f_\Pi} \ , \qquad\qquad\qquad \theta=\frac{v_{S_1}}{\sqrt{v_H^2+v_{S_1}^2}}\sin\frac{\sqrt{v_H^2+v_{S_1}^2}}{f_\Pi} \ ,
\label{eq-vevg2}
\\
&v_H^2=\langle|H|^2\rangle \ , \qquad\qquad\qquad\qquad\qquad\qquad\qquad v_{S_1}^2=\langle|S_1|^2\rangle \ .
\label{eq-vevg3}
\end{align}
In the following we will assume that $v_{S_1}=0$, otherwise SU(3)$_c$ is spontaneously broken, also in the presentation of numerical results we will only show regions of the parameter space satisfying this condition. For vanishing $v_{S_1}$ the vacuum misalignment of Eq.~(\ref{eq-vevg2}) simplifies to the same expression as in the MCHM:
\begin{equation}
\epsilon=\sin\frac{v_H}{f_\Pi} \ , \qquad\qquad\qquad\qquad \theta=0 \ ,
\label{eq-vev}
\end{equation}
with $\epsilon$ determining the amount of EW symmetry breaking (EWSB).

Let us now consider the embedding of ${\cal O}_f$, as well as the resonances of Eq.~(\ref{eq-pc}), into irreducible representations of SO(11). Instead of the usual embeddings in grand unified theories~\cite{Georgi:1974sy,Georgi:1974yf}, we choose representations that satisfy the following conditions: when decomposed under G$_{\rm SM}$ they contain components transforming as the SM fermions, there are SO(10) symmetric Yukawa couplings that contain the usual SM Yukawa interactions with the Higgs, as well as the necessary Yukawa interactions with the LQ: $S_1\bar q^c l$ and/or $S_1\bar u^c e$, and they do not allow di-quark interactions with $S_1$ that would violate baryon number. We show the SO(11) representations in which we embed ${\cal O}_f$, as well as their decomposition under the sequence of subgroups of Eq.~(\ref{eq-subg}) in table~\ref{t-f}.

\begin{table}[ht]
\centering
\begin{tabular}{|c|c|c|c|c|c|}
\hline\rule{0mm}{5mm}
field & $H_{\rm min}$ & SO(6)$\times$SO(4) & SO(10) & SO(11) \\[5pt]
\hline \rule{0mm}{4mm}
$q$ & $({\bf3},{\bf2},{\bf2})_{2}$ & (\bf6,\bf2,\bf2) & {\bf45} & {\bf55} \\[3pt]
\hline  \rule{0mm}{4mm}
$\ell$ & $({\bf1},{\bf2},{\bf2})_{0}$ & $({\bf1},{\bf2},{\bf2})$ & {\bf10} & {\bf11} \\[3pt]
\hline  \rule{0mm}{4mm}
$u$ & $({\bf3},{\bf1},{\bf3})_{2}+({\bf3},{\bf1},{\bf1})_{-4}$ & (\bf6,\bf1,\bf3)+(\bf15,\bf1,\bf1) & {\bf120}+{\bf45} & {\bf165} \\[3pt]
\hline  \rule{0mm}{4mm}
$d$ & $({\bf3},{\bf1},{\bf1})_{2}$ & (\bf6,\bf1,\bf1) & {\bf10} & {\bf11} \\[3pt]
\hline  \rule{0mm}{4mm}
$e$ & $({\bf1},{\bf1},{\bf3})_{0}$ & (\bf1,\bf1,\bf3) & ${\bf45}$ & {\bf55} \\[3pt]
\hline 
\end{tabular}
\caption{Components of the composite partners of the elementary fermions, from the embedding into multiplets of $H_{\rm min}$ up to SO(11).}
\label{t-f}
\end{table}

The representation {\bf165} contains two components transforming under G$_{\rm SM}$ as the singlet quark $u$, as shown in table~\ref{t-f}. Since there is a singlet of SO(10) in ${\bf45\times\bf10\times120}$, and there is no singlet in ${\bf45\times\bf10\times45}$, to leading order in powers of the NGBs only the first component is involved in the Yukawa interaction.

\subsubsection{Conservation of baryon number} 
It is well known that LQs can generically induce baryon decay, requiring their masses to be at least of ${\cal O}(10^{16})$~GeV. Since in the present model the LQ $S_1$ is assumed to be at the TeV scale, and there are also several spin-1 LQs associated to the global currents of SO(11), with masses ${\cal O}(10)$~TeV, these states could interact with di-quarks and generate baryon decay. However the U(1)$_X$ subgroup of SO(11) can be identified with baryon symmetry, with: $B=X/6$, forbidding proton decay. The same subgroup was found to play this role in Ref.~\cite{us}.

Although we are not including a model of neutrino masses in the theory, in the absence of a symmetry for lepton number the Weinberg operator can be generated with a coefficient of order $\epsilon_l^2/m_*$~\cite{1807.04279}, resulting in a contribution to neutrino masses that exceeds the bounds by orders of magnitude. To avoid this effect one can add to the composite sector a global symmetry U(1)$_L$, loosing the original simple group.

\subsection{Flavor structure of the model}
\label{sec-flavor}
We will assume the paradigm of anarchic partial compositeness, that means that there is no structure of flavor in the SCFT and all the flavor transitions are of the same order. At the level of couplings of resonances one gets that for the couplings that are tensors in flavor space, all of their components are of the same order: $c\times g_*$, with $c\sim{\cal O}(1)$. 
Let us consider as an example the interaction between the Higgs boson and composite resonances with the same quantum numbers as a quark doublet and an up-type quark singlet, $Q$ and $U$: 
\begin{equation}
{\cal L}_ {\rm int}^{cp} \supset g_{*H,j\alpha}  \bar Q^c_{Lj} \tilde H U_{R\alpha} \ .
\label{eq-iS2}
\end{equation}
The indices $j$ and $\alpha$ run over three generations, and the coupling $g_{*H,j\alpha}= g_* c_{H,j\alpha}$, with the size of the couplings determined by $1<g_*\ll 4\pi$ and the anarchic flavor structure codified in the uncorrelated numbers $c_{H,j\alpha}\sim{\cal O}(1)$. In sec.~\ref{sec-res} we show how to include this interaction in an SO(11)/SO(10) model of resonances.

The elementary-composite mixing is driven by the coupling $\lambda_f$ of Eq.~(\ref{eq-pc}). Let us consider first the simple case of one generation, and let us rotate away the mixing by performing a unitary rotation of angle $\theta_f$: 
\begin{equation}
f\to f \cos\theta_f-F\sin\theta_f \ , \qquad F\to f \sin\theta_f+F\cos\theta_f \ .
\label{eq-rotf} 
\end{equation}
with the $\theta_f$ and the degree of compositeness $\epsilon_f$ defined by: 
\begin{equation}
\tan\theta_f=\lambda_f/g_* \ , \qquad \epsilon_f\equiv\sin\theta_f \ . 
\label{eq-eps} 
\end{equation}	
After this diagonalization, and before the Higgs acquires a vev, there is one massless chiral fermion with degree of compositeness $\epsilon_f$, and one massive state whose mass is rescaled by the mixing as: $m_*/\cos\theta_f$. 

Fermions with $\epsilon_f\sim 1$ have a large degree of compositeness, as required for the third generation of quarks if the top mass arises from APC, whereas fermions with $\epsilon_f\ll 1$ have a small degree of compositeness and are mostly elementary, as usually required for the first and second generations. 
With the addition of flavor to the model, $\lambda_f$ acquires flavor indices, that can be diagonalized by rotating the elementary and composite fields independently~\cite{Panico-Pomarol,1807.04279}, we will assume this to be the case and take diagonal $\lambda_f$ and $\epsilon_f$.

After the rotation of Eq.~(\ref{eq-eps}), the partially composite fermions obtain Yukawa couplings of order: $\epsilon_{f_L} g_*\epsilon_{f_R}$. 
If one makes the assumption that all the SM fermions acquire mass by APC, the degree of compositeness of most of the fermions is fixed, allowing to estimate the size of their couplings with the SCFT resonances. 
An explanation of the anomalies in $\rd$ requires a large degree of compositeness of the third generation of quarks and leptons, as well as some amount of compositeness of the charm quark. Yet, $\rd$ does not require the rest of the SM fermions to be partially composite, and their masses could arise from an interplay between linear and bilinear interactions~\cite{Panico-Pomarol,DaRold:2017xdm}. This is a very interesting possibility, since it allows to avoid very stringent constraints from some low energy flavor observables involving the fist generation, that require $f_\Pi\gtrsim 20-40$~TeV, one order of magnitude above the scale that we are assuming. 
Therefore, to be able to estimate the size of the corrections from $S_1$, we will assume usual APC with linear interactions only, and we will comment on the particular cases where APC of first generation is problematic.

The hierarchical structure of the Yukawa couplings of the SM can be obtained in APC by choosing $\epsilon_f$~\cite{Agashe:2004cp}.
In the quark sector the hierarchy of the diagonalization matrices leading to the CKM matrix is driven by the hierarchy of Left-handed mixing: $\epsilon_{qj}$, and the hierarchy of masses is determined by the product $\epsilon_{Lj}\epsilon_{Rj}$~\cite{Huber:2003tu}.
We choose the mixing as:
\begin{align}
\epsilon_{q1}\sim\lambda_C^3\epsilon_{q3} \ ,
\qquad
&\epsilon_{u1}\sim \frac{m_u}{v_{\rm SM}}\frac{1}{\lambda_C^3g_*\epsilon_{q3}} \ ,
\qquad
\epsilon_{d1}\sim \frac{m_d}{v_{\rm SM}}\frac{1}{\lambda_C^3g_*\epsilon_{q3}} \ ,
\qquad
\epsilon_{e1}\sim \frac{m_e}{v_{\rm SM}}\frac{1}{g_*\epsilon_{l 1}} \ ,
\nonumber\\
\epsilon_{q2}\sim\lambda_C^2\epsilon_{q3} \ ,
\qquad
&\epsilon_{u2}\sim \frac{m_c}{v_{\rm SM}}\frac{1}{\lambda_C^2g_*\epsilon_{q3}} \ ,
\qquad
\epsilon_{d2}\sim \frac{m_s}{v_{\rm SM}}\frac{1}{\lambda_C^2g_*\epsilon_{q3}} \ ,
\qquad
\epsilon_{e2}\sim \frac{m_\mu}{v_{\rm SM}}\frac{1}{g_*\epsilon_{l 2}} \ ,
\nonumber\\
&\epsilon_{u3}\sim \frac{m_t}{v_{\rm SM}}\frac{1}{g_*\epsilon_{q3}} \ ,
\qquad\quad
\epsilon_{d3}\sim \frac{m_b}{v_{\rm SM}}\frac{1}{g_*\epsilon_{q3}} \ ,
\qquad\quad \
\epsilon_{e3}\sim \frac{m_\tau}{v_{\rm SM}}\frac{1}{g_*\epsilon_{l 3}} \ ,
\label{eq-apc}
\end{align}
%
where the first column allows to reproduce the CKM matrix and the other ones the masses of the charged fermions of the SM. Neutrino masses and PMNS angles require a model for this sector, and to some extent is independent of the phenomenology that we study, thus we do not specify the mechanism of neutrino masses.

APC determines the size of the couplings of Eq.~(\ref{eq-iS1}). The presence of interactions between $S_1$ and the vector-like composite fermions, $S_1 (g_{*L}\bar Q^c_L \tilde \epsilon L_L + g_{*R} \bar U^c_R E_R)$, generates Eq.~(\ref{eq-iS1}) after diagonalization of the mixing generated by $\lambda_f$, with:
\begin{equation}
x_{L,j\alpha}\sim g_* \epsilon_{qj}\epsilon_{l\alpha}c_{L,j\alpha} \ , 
\qquad
x_{R,j\alpha}\sim g_* \epsilon_{uj}\epsilon_{e\alpha}c_{R,j\alpha} \ , 
\label{eq-xpc}
\end{equation}
where $c_{L,R}$ are tensors in flavor space with all of their components independent and ${\cal O}(1)$. See sec.~\ref{sec-res} for their embedding in a theory of resonances with SO(11)/SO(10) symmetry.

\subsection{Effective theory} 
\label{sec-effth}
Once the pattern of spontaneous breaking of the global symmetry is established, and the representations of the fields are chosen, the effective theory at energies below the scale of resonances can be written by following the CCWZ method~\cite{Callan:1969sn}. To build the Lagrangian it is useful to work with the NGB matrix $\Sigma$ in Eq.~(\ref{eq-U}), that transforms non-linearly under an element ${\cal G}\in$SO(11): $\Sigma\to {\cal G}\Sigma{\cal H}^\dagger$, with ${\cal H}$ an element of the unbroken subgroup SO(10) that depends on ${\cal G}$ and $\Pi$.

The NGB kinetic term can be written in terms of the Maurer-Cartan form: $i\Sigma^\dagger D_\mu \Sigma=e^a_\mu T^a+d^{\hat a}_\mu T^{\hat a}$, with $D_\mu$ the covariant derivative containing the elementary gauge fields, $a$ and $\hat a$ labeling the unbroken and broken generators. Thus ${\cal L}_{\rm eff}\supset (f_\Pi^2/4)d^{\hat a}_\mu d^{\mu\hat a}$.
When the Higgs gets a vev a mass for the EW gauge bosons is generated, with:
\begin{equation}
v_{\rm SM}=\epsilon\ f_\Pi \ ,
\label{eq-vSM}
\end{equation}
with $\epsilon$ defined in Eq.~(\ref{eq-vev}).
For $\epsilon\ll 1$ the EW scale of the SM and the scale of the composite sector are separated, leading to composite resonances significantly heavier than $v_{SM}$. EW precision tests require $\epsilon\lesssim 0.1$~\cite{Grojean:2013qca}.

Whereas the fields of the SCFT, as well as the resonances created by them, transform in complete multiplets of SO(11), the elementary fields transform only under the SM gauge group, included in SO(10). To build the effective theory it is useful to embed the elementary fields in the same SO(11) representations as their composite partners of table~\ref{t-f}, by adding spurious fields that are not dynamical and should be put to zero at the end of the calculations. We will denote as $\psi_f$ the SO(11) multiplets containing an elementary field $f$, with f being a multiplet of G$_{\rm SM}$ only.

Dressing with $\Sigma$ the fields embedded in representations of SO(11), $\psi_f$, it is possible to build SO(11) invariants that look like if they were only SO(10) invariants. Decomposing a representation of SO(11) under SO(10) as ${\bf R}\sim\oplus_j {\bf r}_j$, to quadratic order in the elementary fermions ${\cal L}_{\rm eff}$ is written as:
\begin{align}\label{eq-Leff}
&{\cal L}_{\rm eff}\supset \sum_f Z_f\bar\psi_f\pslash\psi_f+
\sum_{f,f'}\sum_{{\bf r}}\left[\Pi_{ff'}^{{\bf r}}(p)(\bar\psi_f \Sigma)P_{{\bf r}}(\Sigma^\dagger\psi_{f'})+\Pi_{ff'^c}^{{\bf r}}(p)(\bar\psi_f \Sigma)P_{{\bf r}}(\Sigma^\dagger\psi_{f'})^c\right]+{\rm h.c.} \nonumber 
\\
&f,f'=q,u,d,l,e \ ,
\end{align}
where the elementary kinetic terms contain a normalization factor $Z_f$, that in numerical calculations will be taken to unity, 
$f^c$ is the charge conjugate of $f$ and generation indices are understood.
$\Pi_{ff'^{(c)}}^{\bf r}(p)$ are the form factors arising from the integration of the heavy resonances of the composite sector, they depend on momentum and on the parameters of that sector, as couplings, masses and decay constants of the resonances.
$P_{{\bf r}}$ is a projector from the representation of SO(11) of $\psi$, to the representation ${\bf r}$ of SO(10), such that a product as $(\bar\psi_f \Sigma)P_{{\bf r}}(\Sigma^\dagger\psi_{f'})^{(c)}$ is superficially an invariant of SO(10), and thanks to the properties of $\Sigma$ it is also an invariant of SO(11). There is one form factor $\Pi_{ff'^{(c)}}$ for each invariant, we give some examples of the form factors contained in ${\cal L}_{\rm eff}$: for terms $\sim\bar\psi_q\times\psi_q$, since ${\bf55}\sim{\bf45}\oplus{\bf10}$, one has $\Pi_{qq}^{\bf45}$ and $\Pi_{qq}^{\bf10}$; for terms $\sim\bar\psi_q\times\psi_u$, since ${\bf165}\sim{\bf120}\oplus{\bf45}$, there is only one common SO(10) representation contained in these fermions and one has $\Pi_{qu}^{\bf45}$; for $\sim\bar\psi_q\times\psi_l^c$, since ${\bf11}\sim{\bf10}\oplus{\bf1}$, the common SO(10) representation contained in these fermions is ${\bf10}$ and one has $\Pi_{ql^c}^{\bf10}$. By making use of the decompositions of Ap.~\ref{ap-so11} it is straightforward to obtain the values that ${\bf r}$ can take for the different correlators. Notice that, although Eq.~(\ref{eq-Leff}) is quadratic in the fermion fields, it contains interactions with the NGBs, that can be obtained by expanding $U$ in powers of $\Pi$.

${\cal L}_{\rm eff}$ also contains terms with the elementary gauge fields. By adding spurious fields to the gauge fields of the SM group, we can embed them in the adjoint representation of SO(11), hereafter called $a_\mu$, and to quadratic order in them we get~\cite{us}:
\begin{align}
&{\cal L}_{\rm eff}\supset 
\frac{1}{2} P_{\mu\nu} [-Z_a a^\mu p^2a^\nu
+\sum_{\bf r}\Pi_{a}^{\bf r}(a^\mu \Sigma)P_{\bf r}(\Sigma^\dagger a^\nu)] \ , 
\label{eq-lg1}
\end{align}
with $Z_a=1/g_{\rm el}^2$ the elementary kinetic term, the projector $P_{\mu\nu}=\eta_{\mu\nu}-p_\mu p_\nu/p^2$ and $\Pi_{g}^{\bf r}$ being the form factors of the gauge fields.

For several calculations it is useful to consider ${\cal L}_{\rm eff}$ to quadratic order in the elementary fields and to all order in the Higgs vev~\cite{us}:
\begin{align}
{\cal L}_{\rm eff}\supset 
&\sum_{f=u,d,e}\sum_{X=L,R}\bar f_X\pslash(Z_{f_X}+\Pi_{f_X})f_X
+
\sum_{f=u,d,e}(\bar f_LM_ff_R+ {\rm h.c.})
\nonumber \\
&+ \bar \nu_L\pslash(Z_{\nu_L}+\Pi_{\nu})\nu_L
+ \frac{1}{2}\sum_{a=g,w,b}\sum_j a^j_\mu (-Z_{a}p^2+\Pi_{a})a^j_\mu 
\ ,
\label{eq-Leffv}
\end{align}
with $Z_{u_L}=Z_{d_L}=Z_q$ and $Z_{\nu_L}=Z_{e_L}=Z_l$ to leading order.
In the second line $g$, $w$ and $b$ are the elementary gauge fields of G$_{SM}$ with couplings $g_{0s}$, $g_{0w}$ and $g_{0b}$, and $Z_a=g_{0a}^{-2}$. 

The form factors in the EWSB vacuum can be expressed in terms of those of Eq.~(\ref{eq-Leff}) and functions of the vev as:
\begin{align}
&\Pi_{f_L}=\sum_{\bf r} i_{f_L}^{\bf r}\hat\Pi_{q}^{\bf r} \ , \qquad
\Pi_{f_R}=\sum_{\bf r} i_{f_R}^{\bf r}\hat\Pi_{f}^{\bf r} \ , \qquad 
M_f=\sum_{\bf r} j_{f}^{\bf r}\Pi_{f}^{\bf r} \ , \qquad
f=u,d,e \ , \nonumber
\\
&\Pi_{\nu}=\sum_{\bf r} i_{\nu_L}^{\bf r}\hat\Pi_{l}^{\bf r} \ , \qquad
\Pi_{a}=i_a^{\bf r}\Pi_{g}^{\bf r} \ , \qquad a=g,w,b \ .
\label{eq-corr}
\end{align}
where $\pslash \hat\Pi_{f}^{\bf r}=\Pi_{ff}^{\bf r}$, factorizing a $\pslash$ in the kinetic form factors, for simplicity. 

Defining $s_v\equiv\sin(v/f_\Pi)$ and $c_v\equiv \cos(v/f_\Pi)$, the functions $i_{f}^{\bf r}, i_a^{\bf r}$ and $j_{f}^{\bf r}$ are given by: 
\begin{equation}
\renewcommand*{\arraystretch}{2}
\arraycolsep=10pt
\begin{array}{llll}
i_{u_L}^{\bf45}=1 \ , & i_{u_L}^{\bf10}=0  \ , 
&i_{d_L}^{\bf45}=1-s_v^2/2 \ , & i_{d_L}^{\bf10}=s_v^2/2  \ ,
\\
i_{u_R}^{\bf120}=1-s_v^2/2 \ , & i_{u_R}^{\bf45}=s_v^2/2  \ , 
&i_{u_R}'^{\bf120}=s_v^2 \ , & i_{u_R}'^{\bf45}=c_v^2  \ ,
\\
i_{d_R}^{\bf10}=1 \ , & i_{d_R}^{\bf1}=0  \ ,
&&
\\
i_{\nu_L}^{\bf10}=1-s_v^2/2 \ , & i_{\nu_L}^{\bf1}=s_v^2/2  \ ,  
&i_{e_L}^{\bf10}=1 \ , & i_{e_L}^{\bf1}=0  \ ,
\\
i_{e_R}^{\bf10}=1-s_v^2/2 \ , & i_{e_R}^{\bf1}=s_v^2/2  \ ,
&&
\end{array}
\label{eq-inv1}
\end{equation}
and
\begin{equation}
\renewcommand*{\arraystretch}{2}
\arraycolsep=10pt
\begin{array}{llll}
j_{u}^{\bf45}=-s_v/\sqrt{2} \ , & j_{d}^{\bf10}=is_v/\sqrt{2}  \ , & j_{e}^{\bf10}=is_v/\sqrt{2}  \ ,
\\
i_g^{\bf45}=1 \ , & i_g^{\bf10}=0 \ , & i_w^{\bf45}=1-s_v^2/2 \ , & i_w^{\bf10}=s_v^2/2 \ .
\end{array}
\label{eq-inv2}
\end{equation}

The spectrum of states, including the resonances that mix with the elementary fields, can be obtained by computing the equation of motion of each species of fields. For a fermion $f$ one has~\cite{us}:
\begin{equation}
p^2(Z_{f_L}+\Pi_{f_L})(Z_{f_R}+\Pi_{f_R})-|M_f^2|=0 \ .
\label{eq-specf}
\end{equation}
For resonances that do not mix with the elementary fields, $Z_f=0$, the spectrum is given by the poles of Eq.~(\ref{eq-specf}). We will make use of these equations in the next sections to obtain the mass of the top quark, as well as the masses of the lightest resonances.

\subsection{Potential}\label{sec-potential}

The interactions between the elementary and composite sectors explicitly break the global SO(10) symmetry, radiatively inducing a potential for the NGBs, that become pseudo-NGBs (pNGBs). In the fermionic sector, this breaking is driven by $\epsilon_f$, since in the present model only $\epsilon_{q3}$, $\epsilon_{u3}$ and $\epsilon_{l3}$ are ${\cal O}(1)$, the effect of these fermions dominate the potential, and the other fermions will not be taken into account. For simplicity we will omit the generation index in this section, assuming that it always refer to the third generation. For the same reasons we will not consider the effect of U(1)$_Y$ in the potential.

At one loop level the Coleman-Weinberg potential can be written as:
\begin{equation}\label{eq-pCW}
V(\Pi)=\frac{1}{2}\int\frac{d^4p}{(2\pi)^4} [-\log\det\ \frac{{\cal K}_f(\Pi)}{{\cal K}_f(0)} + \log\det\ \frac{{\cal K}_a(\Pi)}{{\cal K}_a(0)}] \ ,
\end{equation}
where ${\cal K}_f(\Pi)$ and ${\cal K}_a(\Pi)$ are the pNGB-dependent matrices of fermions and gauge bosons, respectively, obtained by expressing: ${\cal L}_{\rm eff}=\bar F{\cal K}_f(\Pi)F+A{\cal K}_a(\Pi)A$, with $F$ and $A$ the dynamical fermion and gauge fields of the elementary sector, $F={f,f^c}$, $f=q,u,l$, $A=g,w$. 
The denominators subtract a constant term.

Expanding Eq.~(\ref{eq-pCW}) to ${\cal O}(\Pi^2)$ one can obtain the pNGB masses. For a vacuum preserving color we get:
\begin{align}
m_{S_1}^2 = 2\int \frac{d^4p}{(2\pi)^4}\left[
\frac{\hat\Pi_q^{\bf45}-\hat\Pi_q^{\bf10}}{Z_q+\hat\Pi_q^{\bf45}} + \frac{3}{2}\frac{\hat\Pi_u^{\bf45}-\hat\Pi_u^{\bf120}}{Z_u+\hat\Pi_u^{\bf45}+\hat\Pi_u^{\bf120}} -\frac{3}{4} \frac{(\Pi_{ql}^{\bf10})^2}{p^2(Z_q+\hat\Pi_q^{\bf45})(Z_l+\hat\Pi_l^{\bf10})} + 2\frac{\Pi_A^{\bf10}-\Pi_A^{\bf45}}{Z_Ap^2+\Pi_A^{\bf45}}
\right] \ .
\label{eq-mS1}
\end{align}
For simplicity we do not include the expression of the corrections to $m_{S_1}$ from $v_H$, although it is straightforward to include them from the expansion of the potential.

In a vacuum preserving color the potential determining the Higgs vev is~\cite{us}:
\begin{align}
V(h) = \int \frac{d^4p}{(2\pi)^4}&\left\{\frac{9}{2}\log(-Z_wp^2+\Pi_w)-2\sum_{f=u,d,e,\nu}N_f\log[p^2(Z_{f_L}+\Pi_{f_L})(Z_{f_R}+\Pi_{f_R})-|M_f|^2]\right\} \ ,
\label{eq-vh}
\end{align}
with the correlators defined in the effective Lagrangian of Eq.~(\ref{eq-Leffv}), and expressed in terms of a theory of resonances in Ap.~\ref{ap-potential}. Since we have not considered $\nu_R$, one should take $\Pi_{\nu_R}=M_\nu=0$. The Higgs mass can be obtained from the curvature of the potential in the minimum.

Since the dependence of the potential on the parameters of the theory is in general highly non-trivial, it is useful to consider a simplification that allows to estimate the order of magnitude of the pNGB masses and the Higgs vev. We expand the potential to fourth order in powers of the pNGBs:
\begin{equation}
V(\Pi) = m_H^2 |H|^2 + m_{S_1}^2 |S_1|^2 + \lambda_H |H|^4 + \lambda_{S_1} |S_1|^4 + \lambda_{HS_1} |H|^2|S_1|^2 + {\cal O}({\rm \Pi^4}) \ ,
\label{eq-Vap}
\end{equation}
with the quadratic and quartic coefficients being momentum integrals of the correlators. By NDA the order of magnitude of these coefficients can be estimated as: 
\begin{equation}
m_\Pi^2\sim \epsilon_f^2 m_*^4/(4\pi f_\Pi)^2 \ ,\qquad \lambda_\Pi\sim \epsilon_f^2m_*^4/(4\pi f_\Pi^2)^2 \ ,
\label{eq-ev}
\end{equation}
where for $f$ in general one can take the elementary field with the largest degree of compositeness, usually the top quark.

By making use of Eq.~(\ref{eq-vSM}) in (\ref{eq-ev}), for $m_{S_1}^2>0$ the mass of $S_1$ can be estimated as:
\begin{equation}
m_{S_1}^2\sim \left(\frac{g_*^2\epsilon_f v_{\rm SM}}{4\pi \epsilon}\right)^2 \ .
\label{eq-emS1}
\end{equation}

To estimate $\epsilon$ and $m_h$ we expand $V(h)$ to fourth order in $\sin(h/f_\Pi)$:
\begin{equation}
V(h) \simeq  -\alpha \sin^2\left(\frac{h}{f_\Pi}\right)+\beta\sin^4\left(\frac{h}{f_\Pi}\right) \ ,
\label{eq-Vap1}
\end{equation}
where $\alpha$ and $\beta$ can be explicitly expressed as integrals of the correlators, see Ap.~\ref{ap-potential}.
The order of magnitude of these coefficients can be estimated as: $\alpha\sim N_c\epsilon_f^2g_*^4f^4/(4\pi)^2$ and $\beta\sim N_c\epsilon_f^4g_*^4f^4/(4\pi)^2$.
Evaluating $h$ on its vev one gets $\epsilon^2=\alpha/(2\beta)$, already showing that a small $\epsilon$ requires tuning. Expanding to second order in powers of the physical Higgs field around the minimum of the potential one gets~\cite{DaRold:2020bib}:
\begin{equation}
m_h^2 \sim  \frac{8}{f_\Pi^2}\frac{\alpha}{\beta}(\beta-\alpha) \sim \frac{N_c}{2\pi^2}v_{\rm SM}^2\epsilon_f^4g_*^4 \simeq (150\ {\rm GeV})^2 \left(\frac{\epsilon_f}{0.5}\frac{g_*}{3}\right)^4\ .
\label{eq-emh}
\end{equation}
For the prediction with the full calculation at numerical level see sec.~\ref{sec-num}.

\subsection{Theory of resonances: 2-site model}
\label{sec-res}
In this section we define an effective theory describing the lowest lying level of resonances of the SCFT and their interactions with the elementary sector. For that we consider a two-site effective field theory, with site 0 containing the elementary degrees of freedom and site 1 containing the first level of composite resonances created by the operators of the SCFT.~\footnote{In a warped five dimensional realization site-0 would correspond to the ultraviolet boundary and site-1 to the first level of Kaluza-Klein states~\cite{Contino:2004vy,2-site-sundrum}.} 

Site-0 contains the same fields and symmetries as the SM, except that it does not contain the Higgs. The gauge couplings on site-0 are $g_{0s}$, $g_{0w}$ and $g_{0b}$, as in the previous section. Site 1 has an SO(11) gauge symmetry that allows to describe the spin-1 resonances of the SCFT. We assume that SO(11) is spontaneously broken to SO(10) by the strong dynamics, generating the set of composite NGBs identified with the Higgs and the $S_1$ LQ fields, that can be parameterized by the same matrix $\Sigma$ of Eq.~(\ref{eq-U}). On site-1 there are also Dirac fermion fields transforming under SO(11) with the representations of table~\ref{t-f}, we assume that there is one set of fermions for each generation and that all the generations are in the same representations. Site-1 is characterized by a NGB decay constant $f_*$ and by a set of couplings collectively denoted as $g_*$ and satisfying $1\ll g_*\ll 4\pi$, the masses of the states on site-1 are $m_*\sim g_* f_*$. There is a $\sigma$-model field $\Omega$ that transforms bilinearly under the gauge symmetries of both sites: $\Omega\to {\cal G}_0\Omega{\cal G}_1$, with ${\cal G}_{0,1}$ elements of the gauge symmetry group on each site. $\Omega$ connects both sites realizing partial compositeness, and provides the longitudinal degrees of freedom for the spin-1 resonances that become massive. $\Omega$ is characterized by a decay constant $f_\Omega$, of the same order of magnitude as $f_*$, taken of order few TeV. A detailed description of the two-site model can be found, for example, in Refs.~\cite{DeCurtis,Carena-2014}, with a straightforward extension from the coset SO(5)/SO(4) to SO(11)/SO(10).

For the purpose of this article it is enough to show explicitly the term mixing the fermions of both sites and the Yukawa interactions of site-1. 
We use small (capital) letters for fields on site-0 (site-1), and we embed the fields on site-0 into the multiplets of SO(11) of table~\ref{t-f} by adding non-dynamical spurious components. 
By a gauge transformation it is possible to get $\Omega=1$ and the interesting terms simplify to:
\begin{align}
{\cal L}\supset &f_\Omega \sum_{f=q,u,d,l,e} \lambda_f \bar f F 
+ f_*[y_{u*} (\bar Q_L \Sigma)P_{\bf45}(\Sigma^\dagger U_R)+y_{d*} (\bar Q_L \Sigma)P_{\bf10}(\Sigma^\dagger D_R) + y_{e*} (\bar L_L \Sigma)P_{\bf10}(\Sigma^\dagger E_R)]
\nonumber\\
&+ f_*[y_{ql*} (\bar Q_L \Sigma)^c P_{\bf10}^c(\Sigma^\dagger L_L)+y_{ue*} (\bar U_R \Sigma)^c P_{\bf45}^c(\Sigma^\dagger E_R)] + {\rm h.c.} \ ,
\label{eq-lsites}
\end{align}
where on the first line we show the Yukawa interactions that after mixing generate the leading order interactions of the elementary fermions with the Higgs, and in the second line with $S_1$. We have not included interactions with flipped chiralities to keep the potential finite~\cite{DeCurtis,Carena-2014}, they can be included preserving this property by adding one more site~\cite{1210.7114}. 

The matching with the SM gauge couplings is given by: $g_{\rm SM}^{-2}=g_{0}^{-2}+g_{*}^{-2}$, whereas the decay constant of the physical NGBs is: $f_\Pi^{-2}=f_\Omega^{-2}+f_*^{-2}$.

Integrating the resonances of site-1 at tree level one can obtain the form factors of Eq.~(\ref{eq-Leff}) in the site model. We show our result in Ap.~\ref{ap-potential}.

\section{Low energy flavor observables}
\label{sec-flavorobs}
There are a number of low energy observables related with flavor transitions and violation of flavor universality that receive corrections from the presence of $S_1$. Since the rest of the resonances have masses of ${\cal O}(10)$~TeV, in general their effect on these observables is below the present bounds in APC. An exception are some lepton electromagnetic dipole transitions involving $\mu$ and $e$, which are known to require $f_\Pi\gtrsim 20-40$~TeV in scenarios with APC~\cite{Panico:2015jxa}, demanding departures from this paradigm as we will mention in sec.~\ref{sec-EDM}.

We will express the contributions to the low energy flavor observables in terms of the mass and couplings of $S_1$, and then we will estimate their size by making use of partial compositeness. In doing so we will neglect factors of ${\cal O}(1)$, such that the symbol $\sim$ means that the result gives the right order of magnitude, but corrections from number of ${\cal O}(1)$ are expected.

We find it useful to define the following variable~\cite{DaRold:2020bib}:
\begin{equation}
\delta = \frac{g_*^2 v_{\rm SM}^2}{4 m_{S_1}^2} \ ,
\label{eq-delta}
\end{equation}
since this combination will appear repeatedly in the calculations.


\subsection{$\rd$}
\label{sec-fRD}
The effective Lagrangian for $\rd$, after tree level integration of $S_1$, can be written as~\cite{Angelescu:2018tyl,Angelescu:2021lln,Iguro:2022yzr}:
\begin{equation}
{\cal L}_{\rm eff} = 2\sqrt{2} G_F V_{cb}[(1+g_{V_L}){\cal O}_{V_L}+C_{S_L}{\cal O}_{S_L}+C_T{\cal O}_T]
\label{eq-LRD}
\end{equation}
with the Wilson coefficients at the scale $m_{S_1}$ given by:
\begin{align}
& C_{S_L}(m_{S_1}) = -4 C_T(m_{S_1}) = 
-\frac{1}{4\sqrt{2}G_F V_{cb}} \frac{x_{L,33}x_{R,23}^*}{m_{S_1}^2} \ , 
\label{eq-CSL0}
\\
& g_{V_L}(m_{S_1}) = 
\frac{1}{4\sqrt{2}G_F V_{cb}} \frac{x_{L,33}x_{L,23}^*+V_{cs}/V_{cb}\xs{23} \xs{33}^*}{m_{S_1}^2} \ . 
\label{eq-CVL0}
\end{align}

Partial compositeness gives
\begin{align}
& C_{S_L}(m_{S_1}) \sim c_{L,33}c_{R,33}^*\frac{m_c m_\tau}{m_{S_1}^2}\frac{1}{2\lambda_c^4}\simeq 5\times 10^{-4}\left(\frac{\rm TeV}{m_{S_1}}\right)^2 \ , 
\label{eq-CSL}
\\
& g_{V_L}(m_{S_1}) \sim (|c_{L,33}|^2+c_{L,23}c_{L,33}^*)\delta \epsilon_{q3}^2\epsilon_{l3}^2 \sim 0.05 \left(\frac{g_*}{3}\right)^2\left(\frac{\epsilon_{q3}\epsilon_{l3}}{0.6}\right)^2\left(\frac{\rm TeV}{m_{S_1}}\right)^2\ .
\label{eq-CVL}
\end{align}
In the first estimate we kept the $c_i$ factors, one should take into account that in cases with more than one term, as for $g_{V_L}$, there can be phases between the terms that we have put to one for simplicity. For the last estimates we have evaluated the masses, and for $g_{V_L}$ we have taken reference values for the parameters of the model that will be used along this section.

Ref.~\cite{Iguro:2022yzr} finds several fits with different combinations of operators of Eq.~(\ref{eq-LRD}) that can explain the anomalies in $\rd$ and satisfy the bounds from $B_c\to\tau\nu$ and $J/\psi\tau\nu$. Given the estimates of Eq.~(\ref{eq-CSL}) and~(\ref{eq-CVL}), since $C_{S_L}$ is predicted very small we consider the possibility to explain $\rd$ with the operator ${\cal O}_{V_L}$ only. The fit of Ref.~\cite{Iguro:2022yzr} gives $g_{V_L}({\rm TeV})\simeq 0.071\pm 0.018$, that can be easily achieved with partial compositeness and $m_{S_1}\simeq 1$~TeV, by demanding a degree of compositeness of the third generation doublets $\epsilon_{q_3}\epsilon_{l3}\gtrsim 0.6$. In sec.~\ref{sec-RDn} we will show the numerical predictions for this observable.

\subsection{Constraints}
Let us consider now the constraints from flavor physics.
Besides the observables that we explicitly analyze below, we have also studied the following ones: $Z\to\tau\mu$, $B_c\to\tau\nu$, $B_c\to J/\psi\tau\nu$, BR$(Z\to\tau\mu)$, $\mu\to3e$, finding that the corrections from $S_1$ can be safely neglected, since they are well below the bounds.~\footnote{The estimates for most of them in the present model can be obtained from Ref.~\cite{DaRold:2020bib} by considering the limit of decoupling of $S_3$.}


\subsubsection{$R_{b\to c}^{e/\mu}$}
Integrating-out $S_1$ at tree level generates a contribution to this observable, similar to $\rd$~\cite{Isidori,Crivellin,Blanke:2018yud,Blanke:2019qrx}. As in that case, since APC predicts $g_{V_L}\sim \lambda_C^2 C_{S_L}$, the contribution from ${\cal O}_{V_L}$ dominates, with:
\begin{align}
R_{b\to c}^{e/\mu} - 1 & \simeq 
\frac{v_{\rm SM}^2}{2M_{S_1}^2} \left(x_{L,33}x_{L,32}^* + \frac{V_{cs}}{V_{cb}} x_{L,33} x_{L,22}^*\right) \nn \\
&\sim  2 \d1  \epsilon_{q3}^2 \epsilon_{l3}\epsilon_{l2} \left(c_{L,33}c_{L,32}^* + c_{L,33} c_{L,22}^* \right) 
\sim 2\times 10^{-3} \left(\frac{g_*}{3}\right)^{3/2}\frac{\epsilon_{q3}^2\epsilon_{l3}}{0.6}\left(\frac{\rm TeV}{m_{S_1}}\right)^2 \ ,
\label{eq-Rmue}
\end{align}
where we have considered some reference values for the parameters of the model. In particular we estimate $\epsilon_{l2}\sim 0.02/\sqrt{g_*}$, that arises from the muon mass in Eq.~(\ref{eq-apc}) with $\epsilon_{l2}=\epsilon_{e2}$, a limit that is known to minimize several corrections to muon couplings~\cite{Panico:2015jxa}.

The experimental value is~\cite{PDG}: $R_{b\to c,\rm exp}^{\mu/e}-1=0.00 \pm 0.02$, one order of magnitude above the estimate from Eq.~(\ref{eq-Rmue}).

\subsubsection{$B_{K^{(*)}\nu\nu}$}
This observable is modified by tree level exchange of $S_1$. Following Ref.~\cite{Crivellin}, $B\to K^{(*)}\nu \bar{\nu}$ normalized to the SM can be written as:
\begin{equation}
B_{K^{(*)}\nu \bar{\nu}} = \frac{1}{3}\sum_{\alpha,\beta=1,\dots 3}\frac{|C_L^{\rm SM}\delta_{\alpha\beta}+C_L^{\alpha,\beta}|^2}{|C_L^{\rm SM}|^2}
\end{equation}
with $C_\nu^{\rm SM} = -6.4$, $C_L^{\alpha\beta}=C_{L,23}^{\alpha\beta}$ and $C_{L,jk}^{\alpha\beta}$ defined by:
\begin{equation}
C_{L,jk}^{\alpha\beta}=\frac{\sqrt{2}\pi}{4G_F\alpha_{\rm em} V_{td_k}V^*_{td_j}}\frac{x_{L,k\alpha}x_{L,j\beta}^*}{m_{S_1}^2} \ .
\label{eq-CLBKnn}
\end{equation}
Using APC we get:
\begin{equation}
C_L^{\alpha\beta}\sim\frac{2\pi}{\alpha_{\rm em}} \d1 \epsilon_{l3}^2 \epsilon_{q3}^2 c_{L,33}c_{L,23}^* \ ,
\end{equation}

By making use of the experimental constraint $B_{K^{(*)}\nu \bar{\nu},\rm exp}< 2.6$~\cite{1702.03224}, we obtain the bound:
\begin{equation}
g_*\lesssim (1.6-2.2)\frac{m_{S_1}}{{\rm TeV}}\frac{0.6}{\epsilon_{q3}\epsilon_{l3}} \ ,
\end{equation}
where the range of values depend on the sign of the correction, thus this observable is saturated by the $S_1$ contribution.

\subsubsection{$\Delta m_{B_s}$}
The operator ${\cal O}^1_{sb}=(\bar s_L\gamma^\mu b_L)(\bar s_L\gamma_\mu b_L)$ is generated at one loop level, with $S_1$ in a box diagram~\cite{Crivellin}, giving the leading correction to $\Delta m_{B_s}$ that can be written as:
\begin{equation}
\frac{\Delta m_{B_s}}{\Delta m^{\rm SM}_{B_s}} = \Big\vert 1 + \frac{C^1_{sb}}{C^{1,\rm SM}_{sb}}\Big\vert
\end{equation} 
with the coefficient
$C_{sb}^{\rm1, SM} = 2.35 (V_{tb} V_{ts} m_W)^2/(8\pi^2v_{\rm SM}^4)$~\cite{1008.1593}
and $C^1_{sb}$ given by:
\begin{equation}
C^1_{sb} \simeq \frac{1}{128 \pi^2} \left(\frac{\xs{23}^* \xs{33}}{m_{S_1}}\right)^2\sim \left(\frac{\lambda_C^2g_*^2\epsilon_{q3}^2\epsilon_{l3}^2 c_{L,23}^{*2}c_{L,33}^2}{64\pi m_{S_1}}\right)^2 \ ,
\end{equation}
where for the r.h.s we have used the estimates of APC.

Assuming maximally CP violating phases, the most stringent constraint on the imaginary part of $C^1_{sb}$ applies, restricting the effects on $\Delta m_{B_s}$ to be below 20\% at 95\%CL~\cite{1302.0661}. We get the bound:
\begin{equation}
g_*\lesssim 3.2\left(\frac{m_{S_1}}{{\rm TeV}}\right)^{1/2}\frac{0.6}{\epsilon_{q3}\epsilon_{l3}} \ ,
\end{equation}
that is saturated for our choice of parameters.

\subsubsection{$Z$ interactions of $\tau$ and $\nu$}
At one loop level $S_1$ modifies the leptonic interactions of the $Z$, with the corrections to $\tau_L$ and $\nu$ being the most dangerous. We have checked that effects on $\tau_R$ are suppressed compared with Left-handed ones. Parameterizing the interactions at zero momentum as: $(\bar f_L \Gamma_f(0)\gamma^\mu f)Z_\mu$, the leading correction to $\tau_L$ is~\cite{Crivellin}:
\begin{align}
&\Gamma_{\tau_L} - \Gamma_{\tau_L}^{\rm{SM}} = \frac{g}{c_W}\frac{N_c m_t^2}{32 \pi^2}\frac{ V_{3k} \xs{k3}^* V_{3l}^* \xs{l3} }{m_{S_1}^2} \left( 1+\log\left(\frac{m_t^2}{m_{S_1}^2}\right)\right) \ .
\label{eq-dgZtauL}
\end{align} 
Using the SM predictions and the measurements of Ref.~\cite{hep-ex/0509008}, as well as the estimates from APC, we get:
\begin{equation}
g_*\lesssim 4.5\frac{m_{S_1}}{{\rm TeV}}\frac{0.6}{\epsilon_{q3}\epsilon_{l3}} \ .
\label{eq-boundgZtauL}
\end{equation}

Corrections to $\nu$ coupling have been computed in Ref~\cite{Arnan:2019olv}:
\begin{align}
\Gamma_{\nu} - \Gamma^{\rm SM}_{\nu} = \frac{N_c}{16\pi^2}
&\left\{|x_{L,33}|^2\left[(g_{u_L}-g_{u_R})\frac{x_t(x_t-1-\log x_t)}{(x_t-1)^2}+\frac{x_Z}{12}F_2^L(x_t))\right]\right.
\nonumber\\
&+\sum_{k=1,2}|x_{L,k3}|^2\frac{x_Z}{3}\left[g_{u_L}\left(\log x_Z-i\pi-\frac{1}{6}\right)+\frac{g_{\ell_L}}{6}\right]
\nonumber\\
&\left.+\sum_{k=1,2,3}|x_{L,k3}|^2\frac{x_Z}{3}\left[g_{d_L}\left(\log x_Z-i\pi-\frac{1}{6}\right)+\frac{g_{\ell_L}}{6}\right]\right\} \ ,
\label{eq-dgZnu}
\end{align} 
where $g_f=T^{3L}_f-Q_f s_W^2$ is the charge of the SM fermions at tree level, $x_F=m_F^2/m_{S_1}^2$ and:
\begin{align}
F_2^L = 
&-g_{u_L}\frac{(x_t-1)(5x_t^2-7x_t+8)-2(x_t^3+2)\log x_t}{(x_t-1)^4}
\nonumber\\
&-g_{u_R}\frac{(x_t-1)(x_t^2-5x_t-2)+6x_t\log x_t}{(x_t-1)^4}
\nonumber\\
&+g_{\ell_L}\frac{(x_t-1)(-11x_t^2+7x_t-2)+6x_t^3\log x_t}{3(x_t-1)^4} \ .
\end{align} 
Measurements of the number of active neutrinos give bounds on $\Gamma_\nu$. Making use of the relation: $N_\nu = 3 + 4 \, \delta \Gamma_\nu$~\cite{Crivellin} and $N_\nu = 2.9963 \pm 0.0074$~\cite{Janot}, as well as the estimates of APC, we get:
\begin{equation}
g_*\lesssim 6\frac{0.6}{\epsilon_{q3}\epsilon_{l3}} \ ,
\label{eq-boundgZnu}
\end{equation}
where we have taken $m_{S_1}=1$~TeV.

\subsubsection{$W$ interactions of leptons}
At one loop level $S_1$ modifies the $W$ couplings to fermions. In the present model the most relevant modification is to leptons of third generation. Following Refs.~\cite{1705.00929,Isidori} we get:
\begin{equation}
\left\vert\frac{g^W_\tau}{g^W_\ell}\right\vert = 1-0.084 \frac{v_{\rm SM}^2}{4 m_{S_1}^2} |x_{L,33}|^2
\sim 1 - 0.084 \, \d1 \, \epsilon_{q3}^2 \epsilon_{l3}^2 |c_{1,33}|^2 \ ,
\end{equation}
where the r.h.s. is the estimate using APC.

From Ref.~\cite{pich} the ratio $|g^W_\tau/g^W_\ell| $ is measured to be $1.0000 \pm 0.0014$, leading to the bound: 
\begin{equation}
g_*\lesssim 1.8\frac{m_{S_1}}{\rm TeV}\frac{0.6}{\epsilon_{q3}\epsilon_{l3}} \ .
\label{eq-boundgW}
\end{equation}

\subsubsection{Decay $\tau\to 3\mu$}
The decay $\tau\to 3\mu$ is induced at one loop level by the flavor violating coupling $Z\mu\tau$ and four-lepton operators~\cite{Isidori,1705.00929}. The BR$(\tau\to 3\mu)$ is given by:
\begin{equation}
{\rm BR}\left(\tau \to 3\mu \right) 
\simeq 6.25 \times 10^{-5} \ \left|\frac{v_{\rm SM}^2 x_{L,33} x_{L,32}^*}{m_{S_1}^2}\right|^2 
\sim 10^{-3} \,  \d1^2 \epsilon_{q3}^4 \epsilon_{l3}^2 \epsilon_{l2}^2 \left|c_{1,33}^* c_{1,32} \right|^2 \ ,
\label{eq-tau3m}
\end{equation}
where the r.h.s. is the estimate obtained using APC.

The experimental bound at 90\% CL is: BR$(\tau\to 3\mu)<1.2\times 10^{-8}$. Taking $\epsilon_{l2}=\epsilon_{e2}$ we get:
\begin{equation}
g_*\lesssim 7\left(\frac{m_{S_1}}{\rm TeV}\frac{0.6}{\epsilon_{q3}^2\epsilon_{l3}}\right)^{2/3} \ .
\label{eq-boundtau3mu}
\end{equation}

\subsubsection{Decay $\ell_i\to\ell_f\gamma$}
\label{sec-EDM}
These decays are induced by the operators ${\cal O}^{L/R}_{\ell_f\ell_i}=(e/16\pi^2)(\bar \ell_f\sigma^{\mu\nu}P_{L/R}\ell_i)F_{\mu\nu}$, as: 
\begin{equation} 
{\rm BR}(\ell_i \to {\ell_f} \gamma ) = \frac{\alpha_{\rm em} m_{\ell_i}^3 \tau_{\ell_i}}{256 \pi^4} \left( |C^L_{\ell_f \ell_i} |^2 + |C^R_{\ell_f \ell_i}|^2\right)
\end{equation}
with $C^{L/R}_{\ell_f \ell_i}$ the Wilson coefficients of the operators.

$S_1$ generates them at one loop level with Wilson coefficients given by~\cite{Crivellin}:
\begin{equation}
C^L_{\ell_f \ell_i} = - \frac{m_{\ell_f} x_{L,3f}^* x_{L,3i} + m_{\ell_i} x_{R,3f}^* x_{R,3i}}{8 m_{S_1}^2} + \frac{m_t x_{R,3f}^* V^*_{3k} x_{L,ki}}{4 m_{S_1}^2}\left[7+4 \log\left(\frac{m_t^2}{m_{S_1}^2}\right)\right]  \ ,
\end{equation}
with $C^R = {C^L}^\dagger$, due to hermiticity.

We consider first $\tau\to\mu\gamma$, that is dominated by the logarithmic terms. Taking $m_{S_1}=1$~TeV in the logarithm, we obtain:
\begin{equation}
{\rm BR}(\tau \to \mu \gamma) \sim 9 \times 10^{-7} \left(\frac{{\rm TeV}}{m_{S_1}}\right)^4 \frac{\epsilon_{l2}^2}{\epsilon_{l3}^2} \left(c_{R,33}^2 c_{L,32}^{*2} + \left(\frac{m_\mu}{m_\tau}\right)^2 \left(\frac{\epsilon_{l3}}{\epsilon_{l2}}\right)^4c_{R,32}^{*2} c_{L,33}^2\right) \ ,
\label{eq-taumua}
\end{equation}
where we have used APC for the estimate on r.h.s. The experimental bound is ${\rm BR}(\tau \to \mu \gamma)_{\rm exp} < 4.4 \times 10^{-8}$ at 90\%CL~\cite{babar1}.
Eq.~(\ref{eq-taumua}) is minimized for $\epsilon_{l2}\simeq 0.11\epsilon_{l3}$, leading to:
\begin{equation}
m_{S_1}\gtrsim 1.7\ {\rm TeV} \ ,
\label{eq-boundtaumua}
\end{equation}
introducing a slight tension. 

Fixing $\epsilon_{l3}=0.6$, from the minimization of the previous paragraph one gets an $\epsilon_{l2}$ larger than in the LR symmetric limit of the muon, therefore the estimates for a few observables change. The bound from $R^{e/\mu}_{b\to c}$ gives $g_*\lesssim 4$, whereas the one from BR$(\tau\to 3\mu)$ gives $g_*\lesssim 2.3$. 

The BR$(\mu\to e\gamma)$ can be computed following the same procedure. The minimum is found for $\epsilon_{l1}\simeq 0.07\epsilon_{l2}$, that leads to BR$(\mu\to e\gamma)\sim 4\times 10^{-8}({\rm TeV}/m_{S_1})^4$. Since the experimental bound at 90\%CL is ${\rm BR}(\mu \to e \gamma) < 4.2 \times 10^{-13}$~\cite{MEG}, APC requires $m_{S_1}\gtrsim 26$~TeV. Besides this contribution, corrections from regular resonances require $f_\Pi\gtrsim 20$~TeV. An option to evade these stringent constraints is to assume that the electron mass is generated by bilinear interactions, instead of linear ones, allowing to take $\epsilon_{l1}$ and $\epsilon_{e1}$ arbitrarily small, and suppressing dipole operators involving the electron~\cite{DaRold:2017xdm}. See also Refs.~\cite{Matsedonskyi,Panico-Pomarol,DaRold:2021cca} for other scenarios.

\subsubsection{Decays $\tau\to\ell\nu\bar\nu$}
LFU has also been tested in the decays $\tau\to\ell\nu\bar\nu$, that get one loop contributions in the presence of $S_1$~\cite{Feruglio:2016gvd,2008.09548}. By following Refs.~\cite{1310.7922,2008.09548} the strongest constraints arise from the following ratios:
\begin{align}
&\left|\frac{g_\tau}{g_\mu}\right|^2\equiv\frac{\Gamma(\tau\to e\nu\bar\nu)}{\Gamma(\mu\to e\nu\bar\nu)}\left(\frac{\Gamma_{\rm SM}(\tau\to e\nu\bar\nu)}{\Gamma_{\rm SM}(\mu\to e\nu\bar\nu)}\right)^{-1} = 1.0022\pm 0.0030 \ , \nonumber
\\
&\left|\frac{g_\tau}{g_e}\right|^2\equiv\frac{\Gamma(\tau\to \mu\nu\bar\nu)}{\Gamma(\mu\to e\nu\bar\nu)}\left(\frac{\Gamma_{\rm SM}(\tau\to \mu\nu\bar\nu)}{\Gamma_{\rm SM}(\mu\to e\nu\bar\nu)}\right)^{-1} = 1.0058\pm 0.0030 \ ,
\nonumber
\\
&\left|\frac{g_\mu}{g_e}\right|^2\equiv\frac{\Gamma(\tau\to \mu\nu\bar\nu)}{\Gamma(\tau\to e\nu\bar\nu)}\left(\frac{\Gamma_{\rm SM}(\tau\to \mu\nu\bar\nu)}{\Gamma_{\rm SM}(\tau\to e\nu\bar\nu)}\right)^{-1} = 1.0036\pm 0.0028 \ .
\end{align}
The corrections from $S_1$ to the last ratio: $|g_\mu/g_e|^2$, in the pesent scenario with much larger couplings to the third generation, cancel to leading order. Since $|g_\tau/g_e|^2$ is already at almost $2\sigma$ from the SM, we will focus on $|g_\tau/g_\mu|^2$.

For the corrections to these ratios in the presence of leptoquarks we follow Ref.~\cite{2003.12525} that has computed the full one loop matching of a model with $S_1$ (and $S_3$) to the dimension six operators of the SM. They receive contributions from the operator $({\cal O}^{(3)}_{l q})_{\alpha\beta j k}=(\bar l_{L\alpha}\sigma^a\gamma^\mu l_{L\beta})(\bar q_{Lj}\sigma^a\gamma_\mu q_{Lk})$, as well as from the operators generated at one loop: $({\cal O}^{(3)}_{Hl})_{\alpha\beta}=(\bar l_{L\alpha}\sigma^a\gamma^\mu l_{L\beta})(H^\dagger i\overleftrightarrow D_\mu^a H)$ and $({\cal O}_{ll})_{\alpha\beta\gamma\delta}=(\bar l_{L\alpha}\gamma^\mu l_{L\beta})(\bar l_{L\alpha}\gamma_\mu l_{L\beta})$. Following Ref.~\cite{2008.09548} and the hierarchy of couplings of our model determined by APC, we obtain:
\begin{align}
\left|\frac{g_\tau}{g_e}\right|^2 -1 & \simeq-10^{-3}\frac{|x_{L,33}|^2}{(m_{S_1}/{\rm TeV})^2}\left(1.61+0.67\log\frac{m_{S_1}^2}{{\rm TeV}^2}\right) \ ,
\label{eq-gtaugmu}
\\
& \sim 0.1\ \delta\ \epsilon_{q3}^2\epsilon_{\ell 3}^2 |c_{L,33}|^2\left(1+0.42\log\frac{m_{S_1}^2}{{\rm TeV}^2}\right) \ .
\end{align}
Taking into account the sign of the correction, up to factors of ${\cal O}(1)$ we get: 
\begin{equation}
g_*\lesssim 1.2\left(\frac{m_{S_1}}{\rm TeV}\frac{0.6}{\epsilon_{q3}^2\epsilon_{l3}}\right) \ ,
\label{eq-boundgtaugmu}
\end{equation}
introducing some tension, since we have assumed $g_*\simeq 3$ in previous estimates.
 
\subsection{Correlations}
\label{sec-corr}
In the previous subsection we have estimated the corrections to several observables discarding possible correlations. As we will show below, there are correlations between the corrections to $R_{D^{(*)}}$, $B_{K^{(*)}\nu\nu}$ and $\tau\to\ell\nu\bar\nu$ , that lead to constraints that are stronger than those obtained.

From Eqs.~(\ref{eq-CVL0}),~(\ref{eq-CLBKnn}) and~(\ref{eq-gtaugmu}) the main dependence of the corrections to these observables is driven by the following coefficients:
\begin{align}
& G_{R_{D^{(*)}}}=\frac{{\rm Re}(|x_{L,33}|^2+x_{L,33}x^*_{L,23}V_{cs}/V_{cb})}{m_{S_1}^2} \ , \nonumber \\
& G_{B\to K^{(*)}\nu \bar{\nu}}=\frac{{\rm Re}(x_{L,33}x^*_{L,23})}{m_{S_1}^2} \ , \nonumber \\
& G_{\tau\to \ell\nu \bar{\nu}}=\frac{|x_{L,33}|^2}{m_{S_1}^2} \ , 
\label{eq-WG}
\end{align}
from which it is straghtforward to obtain the following result:
\begin{equation}
G_{R_{D^{(*)}}}\simeq {\rm Re}\left(\frac{V_{cs}}{V_{cb}}\right)G_{B\to K^{(*)}}+G_{\tau\to \ell\nu \bar{\nu}} \ .
\label{eq-WGr}
\end{equation}
This relation relys on the hierarchy of couplings determined by APC, that selects the dominant contributions to these processes. Unlike the estimates of the previous sections, that were determined only up to factors of ${\cal O}(1)$, Eq.~(\ref{eq-WGr}) does not suffer from such an indeterminacy, giving a robust relation for an $S_1$ leptoquark with couplings determined by APC.

Using Eq.~(\ref{eq-WGr}), the bounds on $B_{K^{(*)}\nu\nu}$ and $\tau\to\ell\nu\bar\nu$ generate a bound on the corrections to $R_{D^{(*)}}$, we get $g_{V_L}\lesssim 0.035$, at $2\sigma$ of the central value of the fit of Ref.~\cite{Iguro:2022yzr} in the presence of one operator only.

\section{Numerical results}\label{sec-num}
In this section we show different predictions of the model at numerical level. Since the potential is dominated by the mixing of the third generation of fermions in the following we will shorten the notation using $s_f\equiv\epsilon_{f3}$, see Eq.~(\ref{eq-eps}), and $y_*\equiv y_{*,33}$, namely: we will drop the generation index of the elementary-composite mixing and of the composite Yukawa couplings, but they refer to parameters involving the third generation. For the prediction of $\rd$, that depends on $\epsilon_{q2}$, using Eq.~(\ref{eq-apc}) we will replace $\epsilon_{q2}\to\lambda_C^2\epsilon_{q3}$. Besides we choose to work with the masses: $\tilde m_*=m_*/\cos\theta_f$, that is the masses of the fermionic resonances after the diagonalization of the elementary-composite mixing, as described in sec.\ref{sec-flavor}. We define $t_w=g_{0w}/g_*$ and $t_s=g_{0s}/g_*$

We have performed a random scan over the parameter space within the following ranges: $0.5\leq s_{q,u}\leq 0.95$, $0.3\leq s_l\leq 0.8$, $0.2\leq t_{w,s}\leq 0.5$, $|y_{*}|<4.5$, $0.5\leq f_\Pi/{\rm TeV}\leq 10$, $2\leq \tilde m_*/{\rm TeV}\leq 40$, taking the same value of $\tilde m_*$ for all the multiplets. Since we have scanned over 10 parameters and for each point we have computed and minimized the potential, to cover the parameter space with a reasonable density of points consumes a significant amount of CPU time, such that our scan is limited by CPU resources.

\begin{figure}[t]
\centering
\includegraphics[width=0.7\textwidth]{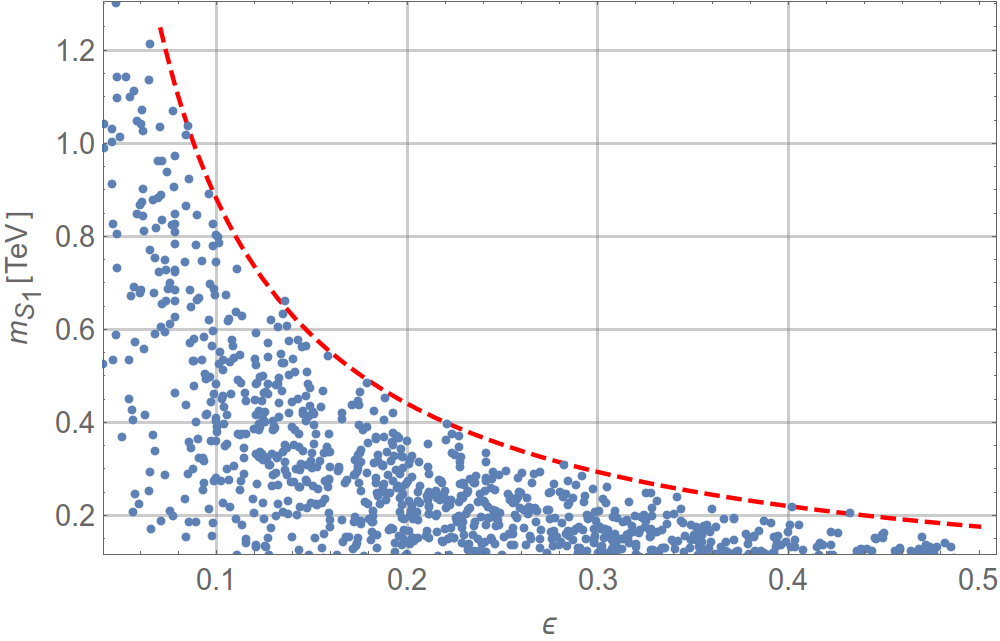}
\caption{$S_1$ mass as function of $\epsilon$ obtained from the one loop potential, with a random scan over the parameters of the model as described in the text. The red line shows the estimate of Eq.~(\ref{eq-emS1}) for the reference values of the parameters. We have normalized dimensionful parameters by fixing $\epsilon f_\Pi=246$~GeV.}
\label{fig-sin-mS1}
\end{figure}

An interesting result of the scan is that we find $m_h\lesssim 140$~GeV and $m_t\lesssim 150$~GeV, notice that these are the masses at a scale of ${\cal O}(10)$~TeV. 
Concerning the LQ mass, in Fig.~\ref{fig-sin-mS1} we show $m_{S_1}$ as function of $\epsilon$ for the random scan. We only show points preserving color symmetry that satisfy: $\epsilon>0$, $f_\Pi>500$~GeV, $m_{S_1}>100$~GeV, $m_h>80$~GeV and $m_t>80$~GeV.
For the plot we have fixed $\epsilon f_\Pi=v_{SM}$, thus $m_{S_1}$ can be estimated by Eq.~(\ref{eq-emS1}) up to factors of ${\cal O}(1)$, the dashed red line shows this estimate with $g_*=3$ and $\epsilon_f=0.5$. As expected $m_{S_1}\gtrsim$~TeV requires $\epsilon\lesssim 0.1$, that for $g_*\gtrsim 3$ gives $m_*\gtrsim{\cal O}(10)$~TeV. Even in this case one can obtain $m_{S_1}<$~TeV if there are partial cancellations between the correlators of the integrand of Eq.~(\ref{eq-mS1}). Thus, one of the main difficulties of the present scenario is the requirement of $m_{S_1}\gtrsim$~TeV, that demands a tuning that has been estimated to be of ${\cal O}(\epsilon^{-2})$~\cite{Agashe-2004,1210.7114} 

From the random scan we have chosen a benchmark point (BP) that has $m_h$ and $m_t$ of ${\cal O}(100)$~GeV and $m_{S_1}\sim$~TeV, with elementary-composite mixing of appropriate size to give $g_{V_L}\sim 0.07$, as discussed in sec.~\ref{sec-fRD}. The BP is defined by:
\begin{align}
& s_q=0.74\ , \qquad\qquad\ \ s_u=0.95\ , \qquad\qquad s_l=0.58\ , \qquad\ \ t_w=0.23\ , \qquad t_s=0.35\ , \nonumber
\\
&y_{u*}=4.24\ , \qquad\qquad\ y_{d*}=0.68, \qquad\qquad y_{e*}=3.90\ ,\qquad y_{ql*}=-2.57\ , \nonumber
\\
&m_*=18.7\ {\rm TeV}\ , \qquad f_\Pi=4.95 \ {\rm TeV}\ . 
\label{eq-BM}
\end{align}
Dimensionful parameters have been rescaled to match the EW scale. Below we will show different predictions using this BP and varying $s_q$ and $s_l$. This BP should be taken as a reference for the phenomenology, taking into account that other points can give slightly different predictions. In particular, one can expect to get BPs with larger $m_{S_1}$ at the price of larger tuning, also in this case larger $y_{ql*}$ will be required to reproduce the anomaly in $\rd$. 

\subsection{Electroweak symmetry breaking (EWSB)}\label{sec-EWSB}
In Fig.~\ref{fig-gvL} the shaded regions show the values of the Higgs vev, according to: light gray for no EWSB: $\epsilon=0$, dark gray for maximal EWSB: $\epsilon=1$, and white for the phenomenological interesting region with $0<\epsilon<1$. The transition from $\epsilon=0$ to $\epsilon=1$ is rather fast, as can be seen from the narrowness of the white region, signaling that points in this region have a large degree of tuning, besides this is the region of phenomenological interest, since for $\epsilon=0$ and $\epsilon=1$ the top quark is massless~\cite{0612048}.
Although not shown in the plot, the line of $\epsilon f_\Pi=246$~GeV is  near and parallel to $\epsilon=0$. Along this line $m_t\sim 50-100$~GeV, and $m_h\simeq m_W$.
The regions of constant $m_t$ are given by lines almost parallel to constant $\epsilon$, in particular the line of $m_t\sim 130$~GeV is very near the line of $\epsilon f_\Pi=246$~GeV
\footnote{One has to consider the running masses at the scale where the resonances are integrated-out, of ${\cal O}(20)\ {\rm TeV}$ for the BP. These masses have to be evolved to lower scales with the RGEs of the SM supplemented with the higher dimensional operators obtained after the integration of the heavy states. We do not attempt to make a precise determination of these masses, we rather check that masses of the proper order of magnitude can be obtained.}. 
In this region also the lines of constant $m_h$ are also almost parallel to lines of constant $\epsilon$. We do not show the lines where these particles have the expected values because the resolution of the plot would not allow a clear distinction between them, as well as with $\epsilon=0$. 

\subsection{$\rd$ versus $R_{K^{(*)}\nu\nu}+|g_\tau/g_\mu|$}\label{sec-RDn}
\begin{figure}[t]
\centering
\includegraphics[width=0.7\textwidth]{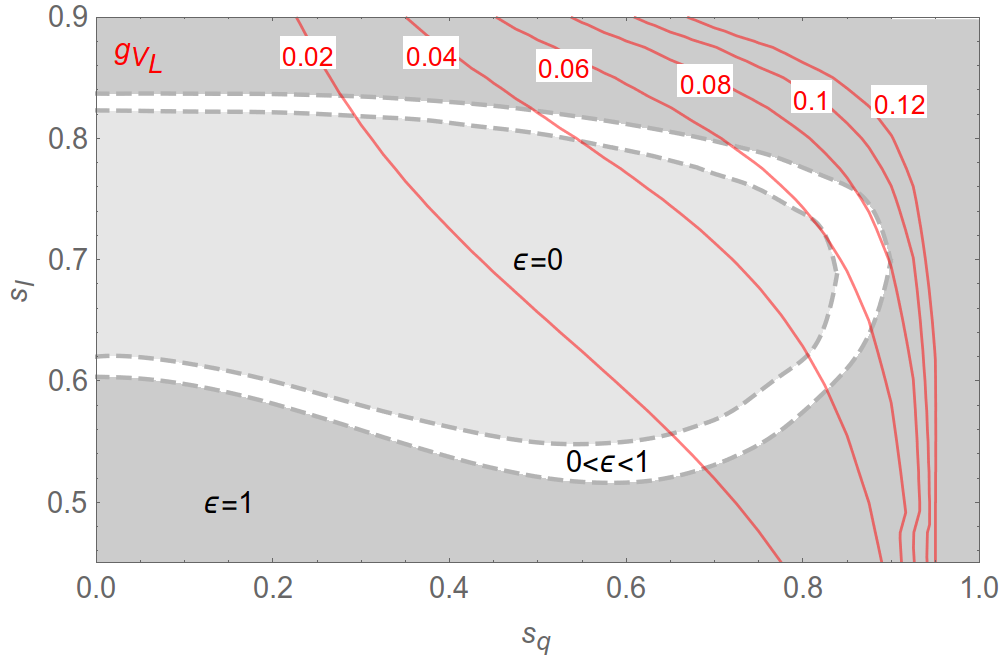}
\caption{In the plane of mixing of the fermion doublets of the third generation, $(s_q,s_l)$, we show in different gray shades the phases of the EW symmetry, with light and dark gray for $\epsilon=0$ and 1, respectively, and white for $0<\epsilon<1$. The red lines show curves of constant $g_{V_L}$, defined in Eq.~(\ref{eq-CVL0}).}
\label{fig-gvL}
\end{figure}

In Fig.~\ref{fig-gvL} we show also the predictions for $g_{V_L}$ in the plane $(s_q,s_l)$, taking the other variables as defined in Eq.~(\ref{eq-BM}) for the BP of the parameter space, the red lines corresponding to constant values of $g_{V_L}$. In this region of the plane $(s_q,s_l)$ the coupling $g_{V_L}$ varies from 0 to $\sim 0.15$, increasing with $s_q$ and $s_l$ as expected from Eq.~(\ref{eq-CVL}), and it is also modulated by $m_{S_1}$, whose values will be shown for this plane in sec.~\ref{sec-resonances}.

\begin{figure}[t]
\centering
\includegraphics[width=0.7\textwidth]{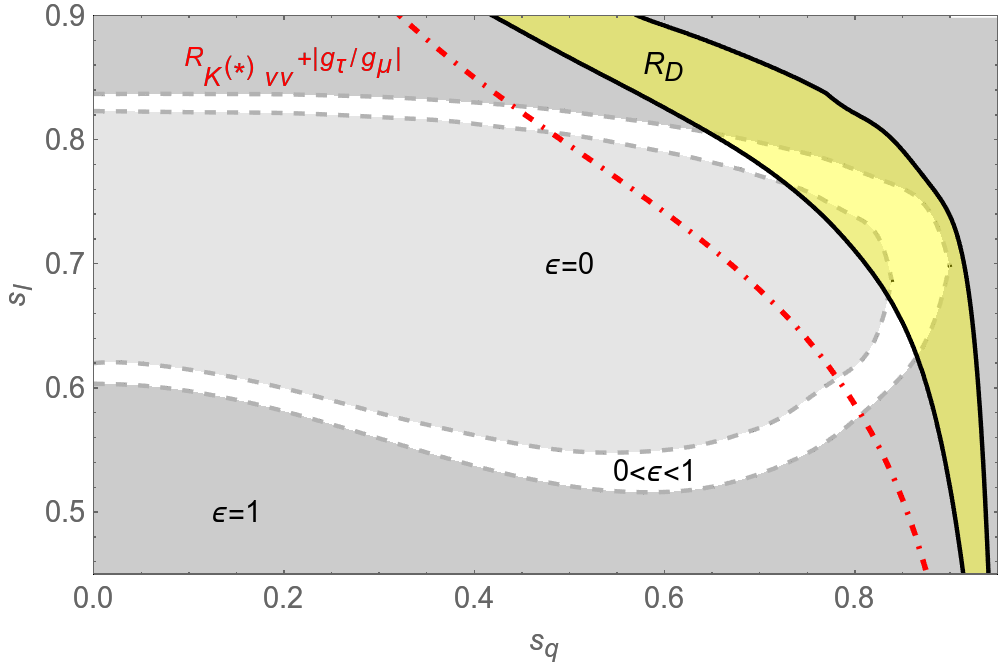}
\caption{In the plane $(s_q,s_l)$ we show in yellow the region where the experimental value of $\rd$ is reproduced at $1\ \sigma$ following the fit of Ref.~\cite{Iguro:2022yzr}, in dot-dashed red the line with constraints from the correlation between $R_{K^{(*)}\nu\nu}+|g_\tau/g_\mu|$, and in different gray shades the phases of the EW symmetry as in Fig.~\ref{fig-gvL}.}
\label{fig-RD}
\end{figure}

In Fig.~\ref{fig-RD} we show in yellow the region where the anomaly in $\rd$ can be reproduced at $1\ \sigma$ following the fit of Ref.~\cite{Iguro:2022yzr}, we show the different phases of the EW symmetry with the same color code as in Fig.~\ref{fig-gvL}, and we also include in dot-dashed red the constraint from $R_{K^{(*)}\nu\nu}+|g_\tau/g_\mu|$, only the region below this line is allowed. Notice that, although a sizable portion of the region with $0<\epsilon<1$ intersects with the region fitting $\rd$, the correlated flavor constraints exclude the one reproducing $\rd$, as discussed in sec.~\ref{sec-corr}. The confirmation of the central value of $\rd$ by future measurements will definitely rule out the model, that can only explain approximately half of that value. 

\section{Phenomenology of the new states}
\label{sec-resonances}
The composite sector is expected to contain towers of resonances transforming in representations of SO(11), with splittings induced by the mixings and the Higgs vev.   
The description in terms of a 2-site model allows to study the phenomenology of the first level of resonances of the composite sector. 

\begin{figure}[t]
\centering
\includegraphics[width=0.7\textwidth]{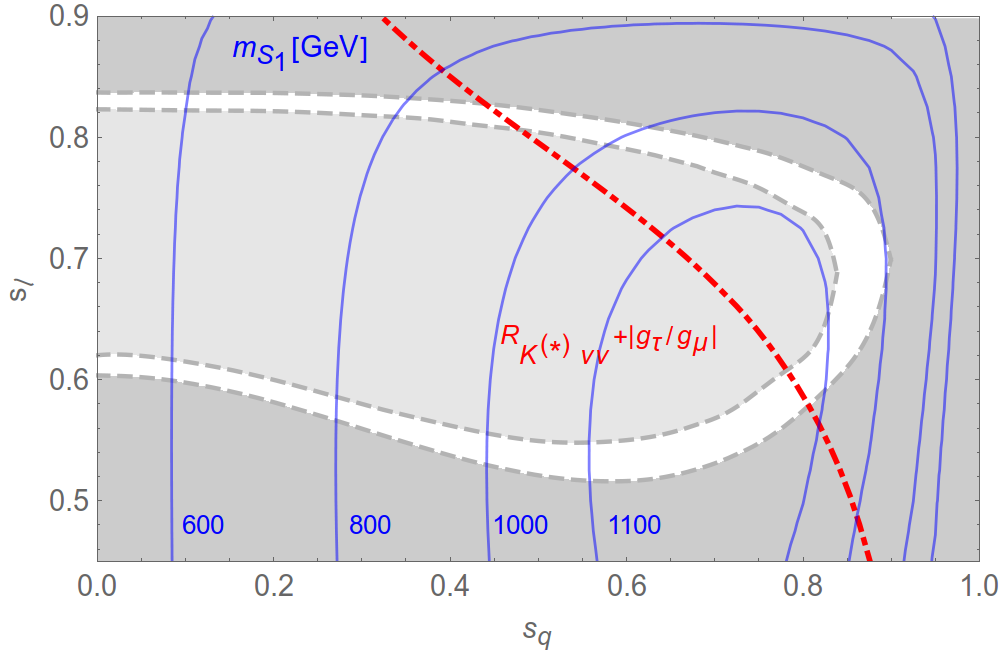}
\caption{In the plane $(s_q,s_l)$ we show in blue the lines of constant $m_{S_1}$, in dot-dashed red the line with constraints from the correlation between $R_{K^{(*)}\nu\nu}+|g_\tau/g_\mu|$, and in different gray shades the phases of the EW symmetry as in Fig.~\ref{fig-gvL}.}
\label{fig-mS1}
\end{figure}

To leading order the mass of $S_1$ can be computed with Eq.~(\ref{eq-mS1}), where corrections from the Higgs vev are not included. In Fig.~\ref{fig-mS1} we show the predictions for the BP in the plane $(s_q,s_l)$. The gray colors show the different phases of the EW symmetry, as in Fig.~\ref{fig-gvL}, and the blue lines show curves of constant $m_{S_1}$. For the region shown $m_{S_1}$ takes values between 0.6 and 1.1 TeV, whereas in the region reproducing the anomaly in $\rd$, $m_{S_1}\sim 1.0-1.1$~TeV.
We have computed the correction to $m_{S_1}$ to ${\cal O}(v_H^2)$, that is suppressed by a factor $\epsilon^2$ compared with the leading order term. Since this correction demands the minimization of the potential, we have only computed it along the curve of $\epsilon f_\Pi\simeq 246$~GeV, obtaining that along the curve the correction can change sign and, since $\epsilon^2\sim 2.5\times 10^{-3}$, it amounts to percent level or less. The maximum value of the corrected mass along that curve is 1.15~TeV. 

$S_1$ decays preferentially to third generation fermions, with a couplings $\sim -1.6$ and BR$(S_1\to b\nu)\simeq$BR$(S_1\to t\tau)$. CMS has considered the double and single production of this state in Ref.~\cite{CMS:2020wzx}, obtaining a bound $m_{S_1}\gtrsim 0.95$~TeV, bounds from ATLAS in $b$-jets plus missing transverse momentum are of order 1~TeV~\cite{ATLAS:2021yij} and in $t\tau$ of order 1.2~TeV~\cite{ATLAS:2021oiz}, introducing some tension with the BP, although the $1\sigma$ limit includes masses up to 1.1 TeV. Therefore, after constraints from direct searches and flavor observables are taken into account, only the region below the dot-dashed red line and the blue line of 1.1 TeV are allowed at $1\sigma$ level. 

The lightest resonances, besides the pNGBs, are the custodians associated to the third generation. Among them, for the BP the lightest state is an up-type quark, $U^{(1)}$, that in the region of the parameter space of fig.~\ref{fig-mS1} has a mass of 4-5 TeV. For $s_{q,l}\to 1$, the mass of the custodians is suppressed as: $m_{\rm cust}=m_*=\tilde m_*\cos\theta_f$, up to corrections from the Higgs vev. Therefore in this limit a set of lepton-like, quark-like and exotic resonances become very light, lighter than $m_{U^{(1)}}$ for $s_f\gtrsim 0.96$. Using the decomposition of the representations ${\bf 11}$, ${\bf 55}$ and ${\bf125}$ given in Ap.~\ref{ap-so11}, one can determine the quantum numbers of these fermions. There are states with the same quantum numbers as the SM fermions, as well as color octets that are either singlets or charged doublets of the EW group, SU(2)$_L$ triplets that are color and hypercharge singlets, and quark-like states with exotic charges: $({\bf3},{\bf2})_{-5/6}$, $({\bf3},{\bf2})_{7/6}$, $({\bf3},{\bf3})_{-1/3}$, $({\bf3},{\bf1})_{-4/3}$. Several of these states have already been found in the MCHM. 

There are also spin-1 resonances that, given that the mixing induced by the gauging is small, have masses of order $m_*\sim 19$~TeV for the BP. These states are associated to the adjoint representation of SO(11). Under G$_{\rm SM}$ there are: colorless states that are triplets of SU(2)$_L$ and have no hypercharge, usually called $W'$, singlets of the EW group, usually called $Z'$, as well as singlets of SU(2)$_L$ with unit electric charge. There is an EW neutral color octet, $G'$, and the following color triplet LQs: $U_1$ and $\bar U_1$, that are SU(2)$_L$ singlets with electric charge 2/3 and -1/3, respectively; as well as $V_2$ and $\tilde V_2$, that are SU(2)$_L$ singlets with electric charge 5/6 and -1/6, respectively, where we have used the notation of Ref.~\cite{Dorsner:2016wpm}.

\section{Conclusions}
\label{sec-conclusions}
It is well known that an $S_1$ LQ with a mass of order TeV, coupled to the third generation and to the quark doublet of the second generation can produce a correction to $\rd$ fitting the experimental measurements. 
In the present paper we have presented a model where $S_1$, as well as the Higgs, are composite NGBs of a new strongly interacting sector. We have shown that if the SCFT has global symmetry SO(11) spontaneously broken down to SO(10) by the strong dynamics, a set of NGBs that matches with $H$ and $S_1$ emerges. We have determined the representations of the operators of the SCFT that contain components with the same quantum numbers as the SM fermions, and that allow the proper Yukawa interactions with Higgs and with $S_1$. We have shown that there is also a conserved U(1) symmetry that can be identified with baryon number.

Avoiding stringent bounds from flavor physics requires a particular flavor structure of the Yukawa couplings of $S_1$.
We have considered that flavor is realized by anarchic partial compositeness, that provides a rationale for the flavor structure of couplings.  We have computed the one loop Coleman-Weinberg potential~\cite{Coleman:1973jx} that is dominated by the interactions with the top quark and the Left-handed bottom and tau, obtaining the masses of the SM states and $S_1$. We have found that there are regions of the parameter space with EW symmetry braking that preserve the color symmetry. We have also found that $m_{S_1}\gtrsim 1$~TeV requires $f_\Pi\gtrsim 5$~TeV and $m_*\sim 20$~TeV. Since the EW scale is one order of magnitude below $f_\Pi$, the tuning is at least of order $10^{-2}-10^{-3}$.

We have computed the corrections to flavor observables by the presence of $S_1$, making extensive use of anarchic partial compositeness. We have found that, 
under the assumption of anarchic partial compositeness, there is a relation between the main coontributions to $\rd$, $B_{K^{(*)}\nu\nu}$ and $g_\tau/g_\mu$, such that the bounds from the last two observables demand $\rd$ to be below approximately half of the present central value, that is $2\sigma$ below it.
Besides the saturation of the bounds from $B_{K^{(*)}\nu\nu}$ and $g_\tau/g_\mu$, there interesting predictions for other flavor observables: 
$g_\tau^W$ demands $g_*\lesssim 2$ for $m_{S_1}=1$~TeV, to be compared with the typical coupling that we have considered that is $g_*=3$, and  BR$(\tau\to\mu\gamma)$ requires 
$m_{S_1}\gtrsim 1.7$~TeV. The next observable putting a relevant constraint is $\Delta m_{B_s}$, whose bound is saturated for the benchmark values of the parameters. 
Besides, satisfying the bound from $\tau\to\mu\gamma$ requires $\epsilon_{e2}\sim 0.04\epsilon_{l2}$, increasing the degree of compositeness of $l_{2}$ compared with the LR symmetric case, and leading to contributions to $R^{e/\mu}_{b\to c}$ and BR$(\tau\to 3\mu)$ that also saturate the bounds. These constraints can be alleviated if the muon mixing departs from APC.
Since the estimates are approximations up to numbers of ${\cal O}(1)$, this tension does not necessarily imply tuning, but it shows an interesting set of observables where deviations must be found in future experiments with more precise measurements, allowing to test the model.

The flavor violating process $\mu\to e\gamma$ is not directly related with $\rd$, but under the assumption of APC one can estimate its rate. Contributions from $S_1$ and from ordinary resonances require $m_{S_1}\gtrsim 26$~TeV and $f_\Pi\gtrsim 20$~TeV, respectively. Thus in the present model that observable demands a departure from APC, for example with the electron having smaller mixing~\cite{Matsedonskyi,Panico-Pomarol,DaRold:2017xdm,DaRold:2021cca}. Such modification does not change the discussion on the anomalies on $\rd$, neither the phenomenology of resonances that we have presented.

Another hint of new physics related with flavor is the anomalous magnetic moment of the muon, however  in APC it has been shown to require an $m_{S_1}$ below the lower bounds. It would be interesting to find a way to explain both deviations in a theory including a rationale for flavor.

We have not discussed the phenomenology of the composite Higgs. Following, for example, Ref.~\cite{Carena-2014}, the corrections to its couplings can be easily computed once the invariants of Eqs.~(\ref{eq-inv1}) and (\ref{eq-inv2}) are known. We leave this study for a future work.

Recently the Belle II collaboration has reported an excess in $B^+\to K^+\nu\bar\nu$~\cite{BelleII:2023}. Refs.~\cite{Bause:2023mfe,Dreiner:2023cms,He:2023bnk,Chen:2023wpb} have studied this signal in models with new states coupled to quarks and leptons, and~\cite{Bause:2023mfe} has shown that agreement with $B_{K^{(*)}\nu\nu}$ requires new LQs at the TeV scale interacting with leptons of the third generation and quarks of second and third generation. It would be interesting to study this excess in the context of pNGB leptoquarks with anarchic partial compositeness.

\section*{Acknowledgments}
I thank Federico Lamagna for sharing results of previous calculations on the subject, as well as Gudrun Hiller and Hector Gisbert for pointing to me the excess of Ref.~\cite{BelleII:2023} and for discussions on this signal, and the referee for pointing to me constraints that were not taken into account in the first version of this article. I also thank financial support from CONICET Argentina with PIP-11220200101426 and FONCyT with PICT-2018-03682.

\appendix
\section{Representations of SO(11)}
\label{ap-so11}
In this appendix we show some useful results of group SO(11).

We have used the following representations and their decompositions under SO(10):
\begin{align}
& {\bf11}\sim{\bf1}\oplus{\bf10} \ , \nonumber
\\
& {\bf55}\sim{\bf10}\oplus{\bf45} \ , \nonumber
\\
& {\bf165}\sim{\bf45}\oplus{\bf120} \ .
\end{align}
Under SO(6)$\times$SO(4) the SO(10) representations obtained decompose as:
\begin{align}
&{\bf45}\sim ({\bf15},{\bf1},{\bf1})\oplus({\bf1},{\bf3},{\bf1})\oplus({\bf1},{\bf1},{\bf3})\oplus({\bf6},{\bf2},{\bf2})
\ , \nonumber
\\
&{\bf120}\sim ({\bf15},{\bf2},{\bf2})\oplus({\bf10},{\bf1},{\bf1})\oplus(\overline	{\bf10},{\bf1},{\bf1})\oplus({\bf6},{\bf3},{\bf1})\oplus({\bf6},{\bf1},{\bf3})\oplus({\bf1},{\bf2},{\bf2})
\ .
\end{align}
the decomposition of ${\bf10}$ is in Eq.~(\ref{eq-NGBdec}).

Under SU(3)$\times$U(1) the SO(6) representations decompose as:
\begin{align}
&{\bf6}\sim {\bf3}_2\oplus\bar{\bf3}_{-2}
\ , \nonumber
\\
&{\bf10}\sim {\bf1}_6\oplus{\bf3}_{2}\oplus{\bf6}_{-2}
\ , \nonumber
\\
&{\bf15}\sim {\bf8}_0\oplus{\bf3}_{-4}\oplus\bar{\bf4}_{4}
\ ,
\end{align}

A basis of the generators of SO(11) in the fundamental representation, meaning squared matrices of dimension eleven, can be obtained by considering the anti-symmetric imaginary matrices:
\begin{equation}
(T^{jk})_{lm}=i(\delta_{jl}\delta_{km}-\delta_{jm}\delta_{kl}) \ , \qquad j=1,\dots k-1 \ , \ k=2,\dots 11 .
\end{equation}
The vacuum $\phi_0$ in Eq.~(\ref{eq-U}) breaks generators with $k$ or $l=5$, and splits the matrices in blocks that can be naturally associated to the unbroken subgroup.

The adjoint representation, ${\bf55}$, can be obtained by computing the structure constants of the group, by using the fundamental representation. Representation ${\bf165}$ was obtained from the product: ${\bf11}\otimes{\bf55}\sim{\bf11}\oplus{\bf165}\oplus{\bf429}$, and identifying ${\bf165}$ in the decomposition.

\section{Form factors}
\label{ap-potential}
In this appendix we collect some useful results involving the form factors of the model.

The coefficients of Eq.~(\ref{eq-Vap1}) can be expressed in terms of the correlators as: 
\begin{align}
\alpha = \int \frac{d^4p}{(2\pi)^4}&
\left(\frac{9}{4}\frac{\Pi_a^{\bf45}-\Pi_a^{\bf10}}{Z_ap^2+\Pi_a^{\bf45}}+\frac{\hat\Pi_l^{\bf10}-\hat\Pi_l^{\bf1}}{Z_l+\hat\Pi_l^{\bf10}}+N_c\frac{\hat\Pi_q^{\bf45}-\hat\Pi_q^{\bf10}}{Z_q+\hat\Pi_q^{\bf45}}+N_c\frac{\hat\Pi_u^{\bf120}-\hat\Pi_u^{\bf45}}{Z_u+\hat\Pi_u^{\bf120}+\hat\Pi_u^{\bf45}}\right.
\nonumber\\
&\left.
+N_c\frac{|\Pi_{qu}^{\bf45}|^2}{p^2(Z_q+\hat\Pi_q^{\bf45})(Z_u+\hat\Pi_u^{\bf120}+\hat\Pi_u^{\bf45})}
\right) \ ,
\label{eq-alpha}
\\
\beta = \int \frac{d^4p}{(2\pi)^4}&\left[-\frac{9}{16}\left(\frac{\Pi_a^{\bf45}-\Pi_a^{\bf10}}{Z_ap^2+\Pi_a^{\bf45}}\right)^2+\frac{1}{4}\left(\frac{\hat\Pi_l^{\bf10}-\hat\Pi_l^{\bf1}}{Z_l+\hat\Pi_l^{\bf10}}\right)^2+\frac{N_c}{4}\left(\frac{\hat\Pi_q^{\bf45}-\hat\Pi_q^{\bf10}}{Z_q+\hat\Pi_q^{\bf45}}\right)^2 \right. \nonumber
\\
&\left.+\frac{N_c}{4}\left(\frac{\hat\Pi_u^{\bf120}-\hat\Pi_u^{\bf45}}{Z_u+\hat\Pi_u^{\bf120}+\hat\Pi_u^{\bf45}}\right)^2 + \frac{N_c}{4}\frac{|\Pi_{qu}^{\bf45}|^4}{p^4(Z_q+\hat\Pi_q^{\bf45})^2(Z_u+\hat\Pi_u^{\bf120}+\hat\Pi_u^{\bf45})^2}\right] \ ,
\label{eq-beta}
\end{align}

In order to obtain the form factors from the 2-site model we define the following functions:
\begin{align}
&A(m_1,m_2,m_3,m_4,y_1,y_2,y_3,\lambda) = \lambda^2 [(m_1^2-p^2+f_*^2y_1^2)((m_2^2-p^2)(m_3^2-p^2)-p^2 f_*^2y_2^2)
\nonumber\\& \qquad\qquad\qquad\qquad\qquad\qquad\qquad\qquad+4(m_3^2-p^2)(m_4^2-p^2)f_*^2y_3^2] \ ,
\nonumber\\
&B(m_1,m_2,m_3,m_4,y_1,y_2,\lambda_1,\lambda_2)= \lambda_1\lambda_2 m_1 m_2 f_* y_1 [(m_3^2-p^2)(m_4^2-p^2)-p^2 f_*^2 y_2^2] \ ,
\nonumber\\
&D(m_1,m_2,m_3,m_4,y_1,y_2,y_3) = 2[(m1^2-p^2)(m_2^2-p^2)-p^2 f_*^2 y_1^2][(m_3^2-p^2)(m_4^2-p^2)-p^2 f_*^2 y_2^2]
\nonumber\\&\qquad\qquad\qquad\qquad\qquad\qquad\qquad-8p^2(m_2^2-p^2)(m_4^2-p^2)f_*^2y_3^2 \ .
\label{eq-auxf}
\end{align}

Using Eq.~(\ref{eq-auxf}) the form factors of Eq.~(\ref{eq-Leff}) can be written as:
\begin{align}
& \hat\Pi_q^{\bf45} = \frac{A(m_u,0,0,0,y_u,0,0,\lambda_q)}{D(m_q,m_u,0,0,y_u,0,0)} \ ,
\qquad\qquad\qquad\
\hat\Pi_q^{\bf10} = \frac{A(m_d,m_l,m_e,m_d,y_d,y_e,y_{ql},\lambda_q)}{D(m_l,m_e,m_q,m_d,y_e,y_d,y_{ql})}\ ,
\nonumber\\
& \hat\Pi_u^{\bf120} = \frac{A(m_q,0,0,0,y_u,0,0,\lambda_u)}{D(m_q,m_u,0,0,y_u,0,0)}\ ,
\qquad\qquad\qquad
\hat\Pi_u^{\bf45} = \frac{A(0,0,0,0,0,0,0,\lambda_u)}{D(m_u,0,0,0,0,0,0)}\ ,
\nonumber\\
&\hat\Pi_d^{\bf10} = \frac{A(m_q,m_l,m_e,0,y_d,y_e,y_{ql},\lambda_d)}{D(m_l,m_e,m_q,m_d,y_e,y_d,y_{ql})}\ ,
\qquad\qquad\ 
\hat\Pi_d^{\bf1} = 2\frac{A(0,0,0,0,0,0,0,\lambda_d)}{D(m_d,0,0,0,0,0,0)} \ ,
\nonumber\\
&\hat\Pi_l^{\bf10} = \frac{A(m_e,0,0,0,y_e,0,0,\lambda_l)}{D(m_l,m_e,0,0,y_e,0,0)}\ ,
\qquad\qquad\qquad\ \ \
\hat\Pi_l^{\bf1} = \frac{A(0,0,0,0,0,0,0,\lambda_l)}{D(m_l,0,0,0,0,0,0)} \ ,
\nonumber\\
&\hat\Pi_e^{\bf45} = \frac{A(0,0,0,0,0,0,0,\lambda_e)}{D(m_e,0,0,0,0,0,0)} \ ,
\qquad\qquad\qquad\qquad
\hat\Pi_e^{\bf10} = \frac{A(m_l,0,0,0,y_e,0,0,\lambda_e)}{D(m_l,m_e,0,0,y_e,0,0)}\ ,
\label{eq-ff}
\end{align}
and
\begin{align}
&\Pi_{qu}^{\bf45} = \frac{B(m_q,m_u,0,0,y_u,0,\lambda_q,\lambda_u)}{D(m_q,m_u,0,0,y_u,0,0)} \ ,
\qquad\qquad\ \ \ \ 
\Pi_{qd}^{\bf10} = \frac{B(m_q,m_d,m_l,m_e,y_d,y_e,\lambda_q,\lambda_d)}{D(m_l,m_e,m_q,m_d,y_e,y_d,y_{ql})}\ ,
\nonumber\\
&\Pi_{le}^{\bf10} = \frac{B(m_l,m_e,0,0,y_e,0,\lambda_l,\lambda_e)}{D(m_l,m_e,0,0,y_e,0,0)}\ ,
\qquad\qquad\qquad
\Pi_{ql}^{\bf10} = 2\frac{B(m_q,m_l,m_d,m_e,y_{ql},0,\lambda_q,\lambda_l)}{D(m_l,m_e,m_q,m_d,y_e,y_d,0)} \ ,
\label{eq-ff1}
\end{align}
For simplicity, and because $\epsilon_e\ll 1$ for all the generations, we have not included the effect of the Yukawa coupling $y_{ue}$ in Eq.~(\ref{eq-ff}).

\bibliographystyle{JHEP}
\bibliography{biblio}

\providecommand{\href}[2]{#2}\begingroup\raggedright\begin{thebibliography}{100}

\bibitem{babar1}
{\scshape BaBar} collaboration, B.~Aubert et~al., \emph{{Searches for Lepton
  Flavor Violation in the Decays tau+- ---\ensuremath{>} e+- gamma and tau+-
  ---\ensuremath{>} mu+- gamma}},
  \href{http://dx.doi.org/10.1103/PhysRevLett.104.021802}{\emph{Phys. Rev.
  Lett.} {\bf 104} (2010) 021802}, [\href{http://arxiv.org/abs/0908.2381}{{\tt
  0908.2381}}].

\bibitem{BaBar:2012obs}
{\scshape BaBar} collaboration, J.~P. Lees et~al., \emph{{Evidence for an
  excess of $\bar{B} \to D^{(*)} \tau^-\bar{\nu}_\tau$ decays}},
  \href{http://dx.doi.org/10.1103/PhysRevLett.109.101802}{\emph{Phys. Rev.
  Lett.} {\bf 109} (2012) 101802}, [\href{http://arxiv.org/abs/1205.5442}{{\tt
  1205.5442}}].

\bibitem{RD_Babar}
{\scshape BaBar} collaboration, J.~Lees et~al., \emph{{Measurement of an Excess
  of $\bar{B} \to D^{(*)}\tau^- \bar{\nu}_\tau$ Decays and Implications for
  Charged Higgs Bosons}},
  \href{http://dx.doi.org/10.1103/PhysRevD.88.072012}{\emph{Phys. Rev. D} {\bf
  88} (2013) 072012}, [\href{http://arxiv.org/abs/1303.0571}{{\tt 1303.0571}}].

\bibitem{Belle:2015qfa}
{\scshape Belle} collaboration, M.~Huschle et~al., \emph{{Measurement of the
  branching ratio of $\bar{B} \to D^{(\ast)} \tau^- \bar{\nu}_\tau$ relative to
  $\bar{B} \to D^{(\ast)} \ell^- \bar{\nu}_\ell$ decays with hadronic tagging
  at Belle}}, \href{http://dx.doi.org/10.1103/PhysRevD.92.072014}{\emph{Phys.
  Rev. D} {\bf 92} (2015) 072014}, [\href{http://arxiv.org/abs/1507.03233}{{\tt
  1507.03233}}].

\bibitem{RD_Belle}
{\scshape Belle} collaboration, S.~Hirose et~al., \emph{{Measurement of the
  $\tau$ lepton polarization and $R(D^*)$ in the decay $\bar{B} \to D^* \tau^-
  \bar{\nu}_\tau$}},
  \href{http://dx.doi.org/10.1103/PhysRevLett.118.211801}{\emph{Phys. Rev.
  Lett.} {\bf 118} (2017) 211801}, [\href{http://arxiv.org/abs/1612.00529}{{\tt
  1612.00529}}].

\bibitem{Belle:2017ilt}
{\scshape Belle} collaboration, S.~Hirose et~al., \emph{{Measurement of the
  $\tau$ lepton polarization and $R(D^*)$ in the decay $\bar{B} \rightarrow D^*
  \tau^- \bar{\nu}_\tau$ with one-prong hadronic $\tau$ decays at Belle}},
  \href{http://dx.doi.org/10.1103/PhysRevD.97.012004}{\emph{Phys. Rev. D} {\bf
  97} (2018) 012004}, [\href{http://arxiv.org/abs/1709.00129}{{\tt
  1709.00129}}].

\bibitem{RD_Belle2}
{\scshape Belle} collaboration, A.~Abdesselam et~al., \emph{{Measurement of
  $\mathcal{R}(D)$ and $\mathcal{R}(D^{\ast})$ with a semileptonic tagging
  method}},  \href{http://arxiv.org/abs/1904.08794}{{\tt 1904.08794}}.

\bibitem{Belle:2019rba}
{\scshape Belle} collaboration, G.~Caria et~al., \emph{{Measurement of
  $\mathcal{R}(D)$ and $\mathcal{R}(D^*)$ with a semileptonic tagging method}},
  \href{http://dx.doi.org/10.1103/PhysRevLett.124.161803}{\emph{Phys. Rev.
  Lett.} {\bf 124} (2020) 161803}, [\href{http://arxiv.org/abs/1910.05864}{{\tt
  1910.05864}}].

\bibitem{RD_LHCB}
{\scshape LHCb} collaboration, R.~Aaij et~al., \emph{{Measurement of the ratio
  of branching fractions $\mathcal{B}(\bar{B}^0 \to
  D^{*+}\tau^{-}\bar{\nu}_{\tau})/\mathcal{B}(\bar{B}^0 \to
  D^{*+}\mu^{-}\bar{\nu}_{\mu})$}},
  \href{http://dx.doi.org/10.1103/PhysRevLett.115.111803}{\emph{Phys. Rev.
  Lett.} {\bf 115} (2015) 111803}, [\href{http://arxiv.org/abs/1506.08614}{{\tt
  1506.08614}}]. [Erratum: Phys.Rev.Lett. 115, 159901 (2015)].

\bibitem{LHCb:2017smo}
{\scshape LHCb} collaboration, R.~Aaij et~al., \emph{{Measurement of the ratio
  of the $B^0 \to D^{*-} \tau^+ \nu_{\tau}$ and $B^0 \to D^{*-} \mu^+
  \nu_{\mu}$ branching fractions using three-prong $\tau$-lepton decays}},
  \href{http://dx.doi.org/10.1103/PhysRevLett.120.171802}{\emph{Phys. Rev.
  Lett.} {\bf 120} (2018) 171802}, [\href{http://arxiv.org/abs/1708.08856}{{\tt
  1708.08856}}].

\bibitem{RD_LHCB2}
{\scshape LHCb} collaboration, R.~Aaij et~al., \emph{{Test of Lepton Flavor
  Universality by the measurement of the $B^0 \to D^{*-} \tau^+ \nu_{\tau}$
  branching fraction using three-prong $\tau$ decays}},
  \href{http://dx.doi.org/10.1103/PhysRevD.97.072013}{\emph{Phys. Rev. D} {\bf
  97} (2018) 072013}, [\href{http://arxiv.org/abs/1711.02505}{{\tt
  1711.02505}}].

\bibitem{London:2021lfn}
D.~London and J.~Matias, \emph{{$B$ Flavour Anomalies: 2021 Theoretical Status
  Report}},
  \href{http://dx.doi.org/10.1146/annurev-nucl-102020-090209}{\emph{Ann. Rev.
  Nucl. Part. Sci.} {\bf 72} (2022) 37--68},
  [\href{http://arxiv.org/abs/2110.13270}{{\tt 2110.13270}}].

\bibitem{LHCb:2022qnv}
{\scshape LHCb} collaboration, R.~Aaij et~al., \emph{{Test of lepton
  universality in $b \rightarrow s \ell^+ \ell^-$ decays}},
  \href{http://dx.doi.org/10.1103/PhysRevLett.131.051803}{\emph{Phys. Rev.
  Lett.} {\bf 131} (2023) 051803}, [\href{http://arxiv.org/abs/2212.09152}{{\tt
  2212.09152}}].

\bibitem{LHCb:2022vje}
{\scshape LHCb} collaboration, R.~Aaij et~al., \emph{{Measurement of lepton
  universality parameters in $B^+\to K^+\ell^+\ell^-$ and $B^0\to
  K^{*0}\ell^+\ell^-$ decays}},
  \href{http://dx.doi.org/10.1103/PhysRevD.108.032002}{\emph{Phys. Rev. D} {\bf
  108} (2023) 032002}, [\href{http://arxiv.org/abs/2212.09153}{{\tt
  2212.09153}}].

\bibitem{EFT_Bhattacharya}
B.~Bhattacharya, A.~Datta, D.~London and S.~Shivashankara, \emph{{Simultaneous
  Explanation of the $R_K$ and $R(D^{(*)})$ Puzzles}},
  \href{http://dx.doi.org/10.1016/j.physletb.2015.02.011}{\emph{Phys. Lett. B}
  {\bf 742} (2015) 370--374}, [\href{http://arxiv.org/abs/1412.7164}{{\tt
  1412.7164}}].

\bibitem{EFT_Azatov}
A.~Azatov, D.~Bardhan, D.~Ghosh, F.~Sgarlata and E.~Venturini, \emph{{Anatomy
  of $b \to c \tau \nu$ anomalies}},
  \href{http://dx.doi.org/10.1007/JHEP11(2018)187}{\emph{JHEP} {\bf 11} (2018)
  187}, [\href{http://arxiv.org/abs/1805.03209}{{\tt 1805.03209}}].

\bibitem{Alonso:2015sja}
R.~Alonso, B.~Grinstein and J.~Martin~Camalich, \emph{{Lepton universality
  violation and lepton flavor conservation in $B$-meson decays}},
  \href{http://dx.doi.org/10.1007/JHEP10(2015)184}{\emph{JHEP} {\bf 10} (2015)
  184}, [\href{http://arxiv.org/abs/1505.05164}{{\tt 1505.05164}}].

\bibitem{EFT_Greljo}
A.~Greljo, G.~Isidori and D.~Marzocca, \emph{{On the breaking of Lepton Flavor
  Universality in B decays}},
  \href{http://dx.doi.org/10.1007/JHEP07(2015)142}{\emph{JHEP} {\bf 07} (2015)
  142}, [\href{http://arxiv.org/abs/1506.01705}{{\tt 1506.01705}}].

\bibitem{EFT_Calibbi}
L.~Calibbi, A.~Crivellin and T.~Ota, \emph{{Effective Field Theory Approach to
  $b\to s\ell\ell^{(')}$, $B\to K^{(*)}\nu\overline{\nu}$ and $B\to
  D^{(*)}\tau\nu$ with Third Generation Couplings}},
  \href{http://dx.doi.org/10.1103/PhysRevLett.115.181801}{\emph{Phys. Rev.
  Lett.} {\bf 115} (2015) 181801}, [\href{http://arxiv.org/abs/1506.02661}{{\tt
  1506.02661}}].

\bibitem{EFT_Bordone}
M.~Bordone, G.~Isidori and S.~Trifinopoulos, \emph{{Semileptonic $B$-physics
  anomalies: A general EFT analysis within $U(2)^n$ flavor symmetry}},
  \href{http://dx.doi.org/10.1103/PhysRevD.96.015038}{\emph{Phys. Rev. D} {\bf
  96} (2017) 015038}, [\href{http://arxiv.org/abs/1702.07238}{{\tt
  1702.07238}}].

\bibitem{Huang:2018nnq}
Z.-R. Huang, Y.~Li, C.-D. Lu, M.~A. Paracha and C.~Wang, \emph{{Footprints of
  New Physics in $b\to c\tau\nu$ Transitions}},
  \href{http://dx.doi.org/10.1103/PhysRevD.98.095018}{\emph{Phys. Rev. D} {\bf
  98} (2018) 095018}, [\href{http://arxiv.org/abs/1808.03565}{{\tt
  1808.03565}}].

\bibitem{Sakaki:2013bfa}
Y.~Sakaki, M.~Tanaka, A.~Tayduganov and R.~Watanabe, \emph{{Testing leptoquark
  models in $\bar B \to D^{(*)} \tau \bar\nu$}},
  \href{http://dx.doi.org/10.1103/PhysRevD.88.094012}{\emph{Phys. Rev. D} {\bf
  88} (2013) 094012}, [\href{http://arxiv.org/abs/1309.0301}{{\tt 1309.0301}}].

\bibitem{Freytsis:2015qca}
M.~Freytsis, Z.~Ligeti and J.~T. Ruderman, \emph{{Flavor models for $\bar{B}
  \to D^{(*)} \tau \bar{\nu}$}},
  \href{http://dx.doi.org/10.1103/PhysRevD.92.054018}{\emph{Phys. Rev. D} {\bf
  92} (2015) 054018}, [\href{http://arxiv.org/abs/1506.08896}{{\tt
  1506.08896}}].

\bibitem{Bauer:2015knc}
M.~Bauer and M.~Neubert, \emph{{Minimal Leptoquark Explanation for the
  $R_{D^{(*)}}$ , $R_K$ , and $(g-2)_\mu$ Anomalies}},
  \href{http://dx.doi.org/10.1103/PhysRevLett.116.141802}{\emph{Phys. Rev.
  Lett.} {\bf 116} (2016) 141802}, [\href{http://arxiv.org/abs/1511.01900}{{\tt
  1511.01900}}].

\bibitem{Li:2016vvp}
X.-Q. Li, Y.-D. Yang and X.~Zhang, \emph{{Revisiting the one leptoquark
  solution to the R(D$^{(*)}$) anomalies and its phenomenological
  implications}}, \href{http://dx.doi.org/10.1007/JHEP08(2016)054}{\emph{JHEP}
  {\bf 08} (2016) 054}, [\href{http://arxiv.org/abs/1605.09308}{{\tt
  1605.09308}}].

\bibitem{Hiller:2016kry}
G.~Hiller, D.~Loose and K.~Sch\"onwald, \emph{{Leptoquark Flavor Patterns \& B
  Decay Anomalies}},
  \href{http://dx.doi.org/10.1007/JHEP12(2016)027}{\emph{JHEP} {\bf 12} (2016)
  027}, [\href{http://arxiv.org/abs/1609.08895}{{\tt 1609.08895}}].

\bibitem{Popov:2016fzr}
O.~Popov and G.~A. White, \emph{{One Leptoquark to unify them? Neutrino masses
  and unification in the light of $(g-2)_\mu$, $R_{D^{(\star)}}$ and $R_K$
  anomalies}},
  \href{http://dx.doi.org/10.1016/j.nuclphysb.2017.08.007}{\emph{Nucl. Phys. B}
  {\bf 923} (2017) 324--338}, [\href{http://arxiv.org/abs/1611.04566}{{\tt
  1611.04566}}].

\bibitem{Cai:2017wry}
Y.~Cai, J.~Gargalionis, M.~A. Schmidt and R.~R. Volkas, \emph{{Reconsidering
  the One Leptoquark solution: flavor anomalies and neutrino mass}},
  \href{http://dx.doi.org/10.1007/JHEP10(2017)047}{\emph{JHEP} {\bf 10} (2017)
  047}, [\href{http://arxiv.org/abs/1704.05849}{{\tt 1704.05849}}].

\bibitem{Isidori}
D.~Buttazzo, A.~Greljo, G.~Isidori and D.~Marzocca, \emph{{B-physics anomalies:
  a guide to combined explanations}},
  \href{http://dx.doi.org/10.1007/JHEP11(2017)044}{\emph{JHEP} {\bf 11} (2017)
  044}, [\href{http://arxiv.org/abs/1706.07808}{{\tt 1706.07808}}].

\bibitem{Angelescu:2018tyl}
A.~Angelescu, D.~Be\v{c}irevi\'c, D.~A. Faroughy and O.~Sumensari,
  \emph{{Closing the window on single leptoquark solutions to the $B$-physics
  anomalies}}, \href{http://dx.doi.org/10.1007/JHEP10(2018)183}{\emph{JHEP}
  {\bf 10} (2018) 183}, [\href{http://arxiv.org/abs/1808.08179}{{\tt
  1808.08179}}].

\bibitem{Iguro:2018vqb}
S.~Iguro, T.~Kitahara, Y.~Omura, R.~Watanabe and K.~Yamamoto, \emph{{D$^{*}$
  polarization vs. $ {R}_{D^{\left(\ast \right)}} $ anomalies in the leptoquark
  models}}, \href{http://dx.doi.org/10.1007/JHEP02(2019)194}{\emph{JHEP} {\bf
  02} (2019) 194}, [\href{http://arxiv.org/abs/1811.08899}{{\tt 1811.08899}}].

\bibitem{Blanke:2018yud}
M.~Blanke, A.~Crivellin, S.~de~Boer, T.~Kitahara, M.~Moscati, U.~Nierste
  et~al., \emph{{Impact of polarization observables and $ B_c\to \tau \nu$ on
  new physics explanations of the $b\to c \tau \nu$ anomaly}},
  \href{http://dx.doi.org/10.1103/PhysRevD.99.075006}{\emph{Phys. Rev. D} {\bf
  99} (2019) 075006}, [\href{http://arxiv.org/abs/1811.09603}{{\tt
  1811.09603}}].

\bibitem{Blanke:2019qrx}
M.~Blanke, A.~Crivellin, T.~Kitahara, M.~Moscati, U.~Nierste and
  I.~Ni\v{s}and\v{z}i\'c, \emph{{Addendum to \textquotedblleft{}Impact of
  polarization observables and $B_c\to \tau \nu$ on new physics explanations of
  the $b\to c \tau \nu$ anomaly''}},
  \href{http://arxiv.org/abs/1905.08253}{{\tt 1905.08253}}. [Addendum:
  Phys.Rev.D 100, 035035 (2019)].

\bibitem{NP_Crivellin2}
A.~Crivellin, C.~Greub, D.~M\"uller and F.~Saturnino, \emph{{Scalar Leptoquarks
  in Leptonic Processes}},  \href{http://arxiv.org/abs/2010.06593}{{\tt
  2010.06593}}.

\bibitem{Bigaran:2019bqv}
I.~Bigaran, J.~Gargalionis and R.~R. Volkas, \emph{{A near-minimal leptoquark
  model for reconciling flavour anomalies and generating radiative neutrino
  masses}}, \href{http://dx.doi.org/10.1007/JHEP10(2019)106}{\emph{JHEP} {\bf
  10} (2019) 106}, [\href{http://arxiv.org/abs/1906.01870}{{\tt 1906.01870}}].

\bibitem{Crivellin}
A.~Crivellin, D.~M\"uller and F.~Saturnino, \emph{{Flavor Phenomenology of the
  Leptoquark Singlet-Triplet Model}},
  \href{http://dx.doi.org/10.1007/JHEP06(2020)020}{\emph{JHEP} {\bf 06} (2020)
  020}, [\href{http://arxiv.org/abs/1912.04224}{{\tt 1912.04224}}].

\bibitem{Cheung:2020sbq}
K.~Cheung, Z.-R. Huang, H.-D. Li, C.-D. L\"u, Y.-N. Mao and R.-Y. Tang,
  \emph{{Revisit to the $b\to c\tau\nu$ transition: In and beyond the SM}},
  \href{http://dx.doi.org/10.1016/j.nuclphysb.2021.115354}{\emph{Nucl. Phys. B}
  {\bf 965} (2021) 115354}, [\href{http://arxiv.org/abs/2002.07272}{{\tt
  2002.07272}}].

\bibitem{Bordone:2020lnb}
M.~Bordone, O.~Cat\`a, T.~Feldmann and R.~Mandal, \emph{{Constraining flavour
  patterns of scalar leptoquarks in the effective field theory}},
  \href{http://dx.doi.org/10.1007/JHEP03(2021)122}{\emph{JHEP} {\bf 03} (2021)
  122}, [\href{http://arxiv.org/abs/2010.03297}{{\tt 2010.03297}}].

\bibitem{Lee:2021jdr}
H.~M. Lee, \emph{{Leptoquark option for B-meson anomalies and leptonic
  signatures}},
  \href{http://dx.doi.org/10.1103/PhysRevD.104.015007}{\emph{Phys. Rev. D} {\bf
  104} (2021) 015007}, [\href{http://arxiv.org/abs/2104.02982}{{\tt
  2104.02982}}].

\bibitem{Angelescu:2021lln}
A.~Angelescu, D.~Be\v{c}irevi\'c, D.~A. Faroughy, F.~Jaffredo and O.~Sumensari,
  \emph{{Single leptoquark solutions to the B-physics anomalies}},
  \href{http://dx.doi.org/10.1103/PhysRevD.104.055017}{\emph{Phys. Rev. D} {\bf
  104} (2021) 055017}, [\href{http://arxiv.org/abs/2103.12504}{{\tt
  2103.12504}}].

\bibitem{Marzocca:2021miv}
D.~Marzocca, S.~Trifinopoulos and E.~Venturini, \emph{{From B-meson anomalies
  to Kaon physics with scalar leptoquarks}},
  \href{http://dx.doi.org/10.1140/epjc/s10052-022-10271-7}{\emph{Eur. Phys. J.
  C} {\bf 82} (2022) 320}, [\href{http://arxiv.org/abs/2106.15630}{{\tt
  2106.15630}}].

\bibitem{Belanger:2021smw}
G.~Belanger et~al., \emph{{Leptoquark manoeuvres in the dark: a simultaneous
  solution of the dark matter problem and the $ {R}_{D^{\left(\ast \right)}} $
  anomalies}}, \href{http://dx.doi.org/10.1007/JHEP02(2022)042}{\emph{JHEP}
  {\bf 02} (2022) 042}, [\href{http://arxiv.org/abs/2111.08027}{{\tt
  2111.08027}}].

\bibitem{Sahoo:2021vug}
S.~Sahoo, S.~Singirala and R.~Mohanta, \emph{{Dark matter and flavor anomalies
  in the light of vector-like fermions and scalar leptoquark}},
  \href{http://arxiv.org/abs/2112.04382}{{\tt 2112.04382}}.

\bibitem{Bhaskar:2022vgk}
A.~Bhaskar, A.~A. Madathil, T.~Mandal and S.~Mitra, \emph{{Combined explanation
  of W-mass, muon g-2, RK(*) and RD(*) anomalies in a singlet-triplet scalar
  leptoquark model}},
  \href{http://dx.doi.org/10.1103/PhysRevD.106.115009}{\emph{Phys. Rev. D} {\bf
  106} (2022) 115009}, [\href{http://arxiv.org/abs/2204.09031}{{\tt
  2204.09031}}].

\bibitem{Li:2022chc}
S.-P. Li, X.-Q. Li, X.-S. Yan and Y.-D. Yang, \emph{{Scotogenic Dirac neutrino
  mass models embedded with leptoquarks: one pathway to address the flavor
  anomalies and the neutrino masses together}},
  \href{http://dx.doi.org/10.1140/epjc/s10052-022-11054-w}{\emph{Eur. Phys. J.
  C} {\bf 82} (2022) 1078}, [\href{http://arxiv.org/abs/2204.09201}{{\tt
  2204.09201}}].

\bibitem{Chen:2022hle}
S.-L. Chen, W.-w. Jiang and Z.-K. Liu, \emph{{Combined explanations of
  B-physics anomalies, $(g-2)_{e, \mu }$ and neutrino masses by scalar
  leptoquarks}},
  \href{http://dx.doi.org/10.1140/epjc/s10052-022-10920-x}{\emph{Eur. Phys. J.
  C} {\bf 82} (2022) 959}, [\href{http://arxiv.org/abs/2205.15794}{{\tt
  2205.15794}}].

\bibitem{Lizana:2023kei}
J.~M. Lizana, J.~Matias and B.~A. Stefanek, \emph{{Explaining the $
  {B}_{d,s}\to {K}^{\left(\ast \right)}{\overline{K}}^{\left(\ast \right)} $
  non-leptonic puzzle and charged-current B-anomalies via scalar leptoquarks}},
  \href{http://dx.doi.org/10.1007/JHEP09(2023)114}{\emph{JHEP} {\bf 09} (2023)
  114}, [\href{http://arxiv.org/abs/2306.09178}{{\tt 2306.09178}}].

\bibitem{2008.09548}
V.~Gherardi, D.~Marzocca and E.~Venturini, \emph{{Low-energy phenomenology of
  scalar leptoquarks at one-loop accuracy}},
  \href{http://dx.doi.org/10.1007/JHEP01(2021)138}{\emph{JHEP} {\bf 01} (2021)
  138}, [\href{http://arxiv.org/abs/2008.09548}{{\tt 2008.09548}}].

\bibitem{Iguro:2022yzr}
S.~Iguro, T.~Kitahara and R.~Watanabe, \emph{{Global fit to $b \to c\tau\nu$
  anomalies 2022 mid-autumn}},  \href{http://arxiv.org/abs/2210.10751}{{\tt
  2210.10751}}.

\bibitem{Gripaios:2014tna}
B.~Gripaios, M.~Nardecchia and S.~A. Renner, \emph{{Composite leptoquarks and
  anomalies in $B$-meson decays}},
  \href{http://dx.doi.org/10.1007/JHEP05(2015)006}{\emph{JHEP} {\bf 05} (2015)
  006}, [\href{http://arxiv.org/abs/1412.1791}{{\tt 1412.1791}}].

\bibitem{Alvarez:2018gxs}
E.~Alvarez, L.~Da~Rold, A.~Juste, M.~Szewc and T.~Vazquez~Schroeder, \emph{{A
  composite pNGB leptoquark at the LHC}},
  \href{http://dx.doi.org/10.1007/JHEP12(2018)027}{\emph{JHEP} {\bf 12} (2018)
  027}, [\href{http://arxiv.org/abs/1808.02063}{{\tt 1808.02063}}].

\bibitem{DaRold:2018moy}
L.~Da~Rold and F.~Lamagna, \emph{{Composite Higgs and leptoquarks from a simple
  group}}, \href{http://dx.doi.org/10.1007/JHEP03(2019)135}{\emph{JHEP} {\bf
  03} (2019) 135}, [\href{http://arxiv.org/abs/1812.08678}{{\tt 1812.08678}}].

\bibitem{DaRold:2020bib}
L.~Da~Rold and F.~Lamagna, \emph{{Model for the singlet-triplet leptoquarks}},
  \href{http://dx.doi.org/10.1103/PhysRevD.103.115007}{\emph{Phys. Rev. D} {\bf
  103} (2021) 115007}, [\href{http://arxiv.org/abs/2011.10061}{{\tt
  2011.10061}}].

\bibitem{Bauer:2009cf}
M.~Bauer, S.~Casagrande, U.~Haisch and M.~Neubert, \emph{{Flavor Physics in the
  Randall-Sundrum Model: II. Tree-Level Weak-Interaction Processes}},
  \href{http://dx.doi.org/10.1007/JHEP09(2010)017}{\emph{JHEP} {\bf 09} (2010)
  017}, [\href{http://arxiv.org/abs/0912.1625}{{\tt 0912.1625}}].

\bibitem{Panico:2015jxa}
G.~Panico and A.~Wulzer, \emph{{The Composite Nambu-Goldstone Higgs}},
  vol.~913.
\newblock Springer, 2016,
  \href{http://dx.doi.org/10.1007/978-3-319-22617-0}{10.1007/978-3-319-22617-0}.

\bibitem{NP_Marzocca}
D.~Marzocca, \emph{{Addressing the B-physics anomalies in a fundamental
  Composite Higgs Model}},
  \href{http://dx.doi.org/10.1007/JHEP07(2018)121}{\emph{JHEP} {\bf 07} (2018)
  121}, [\href{http://arxiv.org/abs/1803.10972}{{\tt 1803.10972}}].

\bibitem{NP_Becirevic2}
D.~Be\v{c}irevi\'c, I.~Dor\v{s}ner, S.~Fajfer, N.~Ko\v{s}nik, D.~A. Faroughy
  and O.~Sumensari, \emph{{Scalar leptoquarks from grand unified theories to
  accommodate the $B$-physics anomalies}},
  \href{http://dx.doi.org/10.1103/PhysRevD.98.055003}{\emph{Phys. Rev. D} {\bf
  98} (2018) 055003}, [\href{http://arxiv.org/abs/1806.05689}{{\tt
  1806.05689}}].

\bibitem{Agashe-2004}
K.~Agashe, R.~Contino and A.~Pomarol, \emph{{The Minimal composite Higgs
  model}}, \href{http://dx.doi.org/10.1016/j.nuclphysb.2005.04.035}{\emph{Nucl.
  Phys. B} {\bf 719} (2005) 165--187},
  [\href{http://arxiv.org/abs/hep-ph/0412089}{{\tt hep-ph/0412089}}].

\bibitem{1210.7114}
G.~Panico, M.~Redi, A.~Tesi and A.~Wulzer, \emph{{On the Tuning and the Mass of
  the Composite Higgs}},
  \href{http://dx.doi.org/10.1007/JHEP03(2013)051}{\emph{JHEP} {\bf 03} (2013)
  051}, [\href{http://arxiv.org/abs/1210.7114}{{\tt 1210.7114}}].

\bibitem{Gherghetta:2000qt}
T.~Gherghetta and A.~Pomarol, \emph{{Bulk fields and supersymmetry in a slice
  of AdS}}, \href{http://dx.doi.org/10.1016/S0550-3213(00)00392-8}{\emph{Nucl.
  Phys. B} {\bf 586} (2000) 141--162},
  [\href{http://arxiv.org/abs/hep-ph/0003129}{{\tt hep-ph/0003129}}].

\bibitem{Agashe:2004cp}
K.~Agashe, G.~Perez and A.~Soni, \emph{{Flavor structure of warped extra
  dimension models}},
  \href{http://dx.doi.org/10.1103/PhysRevD.71.016002}{\emph{Phys. Rev. D} {\bf
  71} (2005) 016002}, [\href{http://arxiv.org/abs/hep-ph/0408134}{{\tt
  hep-ph/0408134}}].

\bibitem{Kaplan:1991dc}
D.~B. Kaplan, \emph{{Flavor at SSC energies: A New mechanism for dynamically
  generated fermion masses}},
  \href{http://dx.doi.org/10.1016/S0550-3213(05)80021-5}{\emph{Nucl. Phys. B}
  {\bf 365} (1991) 259--278}.

\bibitem{Contino:2004vy}
R.~Contino and A.~Pomarol, \emph{{Holography for fermions}},
  \href{http://dx.doi.org/10.1088/1126-6708/2004/11/058}{\emph{JHEP} {\bf 11}
  (2004) 058}, [\href{http://arxiv.org/abs/hep-th/0406257}{{\tt
  hep-th/0406257}}].

\bibitem{Frigerio:2011zg}
M.~Frigerio, J.~Serra and A.~Varagnolo, \emph{{Composite GUTs: models and
  expectations at the LHC}},
  \href{http://dx.doi.org/10.1007/JHEP06(2011)029}{\emph{JHEP} {\bf 06} (2011)
  029}, [\href{http://arxiv.org/abs/1103.2997}{{\tt 1103.2997}}].

\bibitem{Hosotani:2015hoa}
Y.~Hosotani and N.~Yamatsu, \emph{{Gauge\textendash{}Higgs grand unification}},
  \href{http://dx.doi.org/10.1093/ptep/ptv153}{\emph{PTEP} {\bf 2015} (2015)
  111B01}, [\href{http://arxiv.org/abs/1504.03817}{{\tt 1504.03817}}].

\bibitem{Aydemir:2018cbb}
U.~Aydemir, D.~Minic, C.~Sun and T.~Takeuchi, \emph{{$B$-decay anomalies and
  scalar leptoquarks in unified Pati-Salam models from noncommutative
  geometry}}, \href{http://dx.doi.org/10.1007/JHEP09(2018)117}{\emph{JHEP} {\bf
  09} (2018) 117}, [\href{http://arxiv.org/abs/1804.05844}{{\tt 1804.05844}}].

\bibitem{Becirevic:2018afm}
D.~Be\v{c}irevi\'c, I.~Dor\v{s}ner, S.~Fajfer, N.~Ko\v{s}nik, D.~A. Faroughy
  and O.~Sumensari, \emph{{Scalar leptoquarks from grand unified theories to
  accommodate the $B$-physics anomalies}},
  \href{http://dx.doi.org/10.1103/PhysRevD.98.055003}{\emph{Phys. Rev. D} {\bf
  98} (2018) 055003}, [\href{http://arxiv.org/abs/1806.05689}{{\tt
  1806.05689}}].

\bibitem{Aydemir:2019ynb}
U.~Aydemir, T.~Mandal and S.~Mitra, \emph{{Addressing the ${\mathbf
  R_{D^{(*)}}}$ anomalies with an ${\mathbf S_1}$ leptoquark from
  $\mathbf{SO(10)}$ grand unification}},
  \href{http://dx.doi.org/10.1103/PhysRevD.101.015011}{\emph{Phys. Rev. D} {\bf
  101} (2020) 015011}, [\href{http://arxiv.org/abs/1902.08108}{{\tt
  1902.08108}}].

\bibitem{Georgi:1974sy}
H.~Georgi and S.~L. Glashow, \emph{{Unity of All Elementary Particle Forces}},
  \href{http://dx.doi.org/10.1103/PhysRevLett.32.438}{\emph{Phys. Rev. Lett.}
  {\bf 32} (1974) 438--441}.

\bibitem{Georgi:1974yf}
H.~Georgi, H.~R. Quinn and S.~Weinberg, \emph{{Hierarchy of Interactions in
  Unified Gauge Theories}},
  \href{http://dx.doi.org/10.1103/PhysRevLett.33.451}{\emph{Phys. Rev. Lett.}
  {\bf 33} (1974) 451--454}.

\bibitem{us}
L.~Da~Rold and F.~Lamagna, \emph{{A vector leptoquark for the B-physics
  anomalies from a composite GUT}},
  \href{http://dx.doi.org/10.1007/JHEP12(2019)112}{\emph{JHEP} {\bf 12} (2019)
  112}, [\href{http://arxiv.org/abs/1906.11666}{{\tt 1906.11666}}].

\bibitem{1807.04279}
M.~Frigerio, M.~Nardecchia, J.~Serra and L.~Vecchi, \emph{{The Bearable
  Compositeness of Leptons}},
  \href{http://dx.doi.org/10.1007/JHEP10(2018)017}{\emph{JHEP} {\bf 10} (2018)
  017}, [\href{http://arxiv.org/abs/1807.04279}{{\tt 1807.04279}}].

\bibitem{Panico-Pomarol}
G.~Panico and A.~Pomarol, \emph{{Flavor hierarchies from dynamical scales}},
  \href{http://dx.doi.org/10.1007/JHEP07(2016)097}{\emph{JHEP} {\bf 07} (2016)
  097}, [\href{http://arxiv.org/abs/1603.06609}{{\tt 1603.06609}}].

\bibitem{DaRold:2017xdm}
L.~Da~Rold, \emph{{Anarchy with linear and bilinear interactions}},
  \href{http://dx.doi.org/10.1007/JHEP10(2017)120}{\emph{JHEP} {\bf 10} (2017)
  120}, [\href{http://arxiv.org/abs/1708.08515}{{\tt 1708.08515}}].

\bibitem{Huber:2003tu}
S.~J. Huber, \emph{{Flavor violation and warped geometry}},
  \href{http://dx.doi.org/10.1016/S0550-3213(03)00502-9}{\emph{Nucl. Phys. B}
  {\bf 666} (2003) 269--288}, [\href{http://arxiv.org/abs/hep-ph/0303183}{{\tt
  hep-ph/0303183}}].

\bibitem{Callan:1969sn}
C.~G. Callan, Jr., S.~R. Coleman, J.~Wess and B.~Zumino, \emph{{Structure of
  phenomenological Lagrangians. 2.}},
  \href{http://dx.doi.org/10.1103/PhysRev.177.2247}{\emph{Phys. Rev.} {\bf 177}
  (1969) 2247--2250}.

\bibitem{Grojean:2013qca}
C.~Grojean, O.~Matsedonskyi and G.~Panico, \emph{{Light top partners and
  precision physics}},
  \href{http://dx.doi.org/10.1007/JHEP10(2013)160}{\emph{JHEP} {\bf 10} (2013)
  160}, [\href{http://arxiv.org/abs/1306.4655}{{\tt 1306.4655}}].

\bibitem{2-site-sundrum}
R.~Contino, T.~Kramer, M.~Son and R.~Sundrum, \emph{{Warped/composite
  phenomenology simplified}},
  \href{http://dx.doi.org/10.1088/1126-6708/2007/05/074}{\emph{JHEP} {\bf 05}
  (2007) 074}, [\href{http://arxiv.org/abs/hep-ph/0612180}{{\tt
  hep-ph/0612180}}].

\bibitem{DeCurtis}
S.~De~Curtis, M.~Redi and A.~Tesi, \emph{{The 4D Composite Higgs}},
  \href{http://dx.doi.org/10.1007/JHEP04(2012)042}{\emph{JHEP} {\bf 04} (2012)
  042}, [\href{http://arxiv.org/abs/1110.1613}{{\tt 1110.1613}}].

\bibitem{Carena-2014}
M.~Carena, L.~Da~Rold and E.~Pont\'on, \emph{{Minimal Composite Higgs Models at
  the LHC}}, \href{http://dx.doi.org/10.1007/JHEP06(2014)159}{\emph{JHEP} {\bf
  06} (2014) 159}, [\href{http://arxiv.org/abs/1402.2987}{{\tt 1402.2987}}].

\bibitem{PDG}
{\scshape Particle Data Group} collaboration, P.~Zyla et~al., \emph{{Review of
  Particle Physics}}, \href{http://dx.doi.org/10.1093/ptep/ptaa104}{\emph{PTEP}
  {\bf 2020} (2020) 083C01}.

\bibitem{1702.03224}
{\scshape Belle} collaboration, J.~Grygier et~al., \emph{{Search for
  $\boldsymbol{B\to h\nu\bar{\nu}}$ decays with semileptonic tagging at
  Belle}}, \href{http://dx.doi.org/10.1103/PhysRevD.96.091101}{\emph{Phys. Rev.
  D} {\bf 96} (2017) 091101}, [\href{http://arxiv.org/abs/1702.03224}{{\tt
  1702.03224}}]. [Addendum: Phys.Rev.D 97, 099902 (2018)].

\bibitem{1008.1593}
A.~Lenz, U.~Nierste, J.~Charles, S.~Descotes-Genon, A.~Jantsch, C.~Kaufhold
  et~al., \emph{{Anatomy of New Physics in $B - \bar{B}$ mixing}},
  \href{http://dx.doi.org/10.1103/PhysRevD.83.036004}{\emph{Phys. Rev. D} {\bf
  83} (2011) 036004}, [\href{http://arxiv.org/abs/1008.1593}{{\tt 1008.1593}}].

\bibitem{1302.0661}
G.~Isidori, \emph{{Flavor physics and CP violation}},  in \emph{{2012 European
  School of High-Energy Physics}}, pp.~69--105, 2014.
\newblock \href{http://arxiv.org/abs/1302.0661}{{\tt 1302.0661}}.
\newblock \href{http://dx.doi.org/10.5170/CERN-2014-008.69}{DOI}.

\bibitem{hep-ex/0509008}
{\scshape ALEPH, DELPHI, L3, OPAL, SLD, LEP Electroweak Working Group, SLD
  Electroweak Group, SLD Heavy Flavour Group} collaboration, S.~Schael et~al.,
  \emph{{Precision electroweak measurements on the $Z$ resonance}},
  \href{http://dx.doi.org/10.1016/j.physrep.2005.12.006}{\emph{Phys. Rept.}
  {\bf 427} (2006) 257--454}, [\href{http://arxiv.org/abs/hep-ex/0509008}{{\tt
  hep-ex/0509008}}].

\bibitem{Arnan:2019olv}
P.~Arnan, D.~Becirevic, F.~Mescia and O.~Sumensari, \emph{{Probing low energy
  scalar leptoquarks by the leptonic $W$ and $Z$ couplings}},
  \href{http://dx.doi.org/10.1007/JHEP02(2019)109}{\emph{JHEP} {\bf 02} (2019)
  109}, [\href{http://arxiv.org/abs/1901.06315}{{\tt 1901.06315}}].

\bibitem{Janot}
P.~Janot and S.~Jadach, \emph{{Improved Bhabha cross section at LEP and the
  number of light neutrino species}},
  \href{http://dx.doi.org/10.1016/j.physletb.2020.135319}{\emph{Phys. Lett. B}
  {\bf 803} (2020) 135319}, [\href{http://arxiv.org/abs/1912.02067}{{\tt
  1912.02067}}].

\bibitem{1705.00929}
F.~Feruglio, P.~Paradisi and A.~Pattori, \emph{{On the Importance of
  Electroweak Corrections for B Anomalies}},
  \href{http://dx.doi.org/10.1007/JHEP09(2017)061}{\emph{JHEP} {\bf 09} (2017)
  061}, [\href{http://arxiv.org/abs/1705.00929}{{\tt 1705.00929}}].

\bibitem{pich}
A.~Pich, \emph{{Precision Tau Physics}},
  \href{http://dx.doi.org/10.1016/j.ppnp.2013.11.002}{\emph{Prog. Part. Nucl.
  Phys.} {\bf 75} (2014) 41--85}, [\href{http://arxiv.org/abs/1310.7922}{{\tt
  1310.7922}}].

\bibitem{MEG}
{\scshape MEG} collaboration, A.~M. Baldini et~al., \emph{{Search for the
  lepton flavour violating decay $\mu ^+ \rightarrow \mathrm {e}^+ \gamma $
  with the full dataset of the MEG experiment}},
  \href{http://dx.doi.org/10.1140/epjc/s10052-016-4271-x}{\emph{Eur. Phys. J.
  C} {\bf 76} (2016) 434}, [\href{http://arxiv.org/abs/1605.05081}{{\tt
  1605.05081}}].

\bibitem{Matsedonskyi}
O.~Matsedonskyi, \emph{{On Flavour and Naturalness of Composite Higgs Models}},
  \href{http://dx.doi.org/10.1007/JHEP02(2015)154}{\emph{JHEP} {\bf 02} (2015)
  154}, [\href{http://arxiv.org/abs/1411.4638}{{\tt 1411.4638}}].

\bibitem{DaRold:2021cca}
L.~Da~Rold and F.~Lamagna, \emph{{Composite Froggatt-Nielsen model of flavor}},
  \href{http://dx.doi.org/10.1103/PhysRevD.105.115020}{\emph{Phys. Rev. D} {\bf
  105} (2022) 115020}, [\href{http://arxiv.org/abs/2112.14600}{{\tt
  2112.14600}}].

\bibitem{Feruglio:2016gvd}
F.~Feruglio, P.~Paradisi and A.~Pattori, \emph{{Revisiting Lepton Flavor
  Universality in B Decays}},
  \href{http://dx.doi.org/10.1103/PhysRevLett.118.011801}{\emph{Phys. Rev.
  Lett.} {\bf 118} (2017) 011801}, [\href{http://arxiv.org/abs/1606.00524}{{\tt
  1606.00524}}].

\bibitem{1310.7922}
A.~Pich, \emph{{Precision Tau Physics}},
  \href{http://dx.doi.org/10.1016/j.ppnp.2013.11.002}{\emph{Prog. Part. Nucl.
  Phys.} {\bf 75} (2014) 41--85}, [\href{http://arxiv.org/abs/1310.7922}{{\tt
  1310.7922}}].

\bibitem{2003.12525}
V.~Gherardi, D.~Marzocca and E.~Venturini, \emph{{Matching scalar leptoquarks
  to the SMEFT at one loop}},
  \href{http://dx.doi.org/10.1007/JHEP07(2020)225}{\emph{JHEP} {\bf 07} (2020)
  225}, [\href{http://arxiv.org/abs/2003.12525}{{\tt 2003.12525}}]. [Erratum:
  JHEP 01, 006 (2021)].

\bibitem{0612048}
R.~Contino, L.~Da~Rold and A.~Pomarol, \emph{{Light custodians in natural
  composite Higgs models}},
  \href{http://dx.doi.org/10.1103/PhysRevD.75.055014}{\emph{Phys. Rev. D} {\bf
  75} (2007) 055014}, [\href{http://arxiv.org/abs/hep-ph/0612048}{{\tt
  hep-ph/0612048}}].

\bibitem{CMS:2020wzx}
{\scshape CMS} collaboration, A.~M. Sirunyan et~al., \emph{{Search for singly
  and pair-produced leptoquarks coupling to third-generation fermions in
  proton-proton collisions at s=13~TeV}},
  \href{http://dx.doi.org/10.1016/j.physletb.2021.136446}{\emph{Phys. Lett. B}
  {\bf 819} (2021) 136446}, [\href{http://arxiv.org/abs/2012.04178}{{\tt
  2012.04178}}].

\bibitem{ATLAS:2021yij}
{\scshape ATLAS} collaboration, G.~Aad et~al., \emph{{Search for new phenomena
  in final states with $b$-jets and missing transverse momentum in
  $\sqrt{s}=13$ TeV $pp$ collisions with the ATLAS detector}},
  \href{http://dx.doi.org/10.1007/JHEP05(2021)093}{\emph{JHEP} {\bf 05} (2021)
  093}, [\href{http://arxiv.org/abs/2101.12527}{{\tt 2101.12527}}].

\bibitem{ATLAS:2021oiz}
{\scshape ATLAS} collaboration, G.~Aad et~al., \emph{{Search for pair
  production of third-generation scalar leptoquarks decaying into a top quark
  and a $\tau$-lepton in $pp$ collisions at $ \sqrt{s} $ = 13 TeV with the
  ATLAS detector}},
  \href{http://dx.doi.org/10.1007/JHEP06(2021)179}{\emph{JHEP} {\bf 06} (2021)
  179}, [\href{http://arxiv.org/abs/2101.11582}{{\tt 2101.11582}}].

\bibitem{Dorsner:2016wpm}
I.~Dor\v{s}ner, S.~Fajfer, A.~Greljo, J.~F. Kamenik and N.~Ko\v{s}nik,
  \emph{{Physics of leptoquarks in precision experiments and at particle
  colliders}},
  \href{http://dx.doi.org/10.1016/j.physrep.2016.06.001}{\emph{Phys. Rept.}
  {\bf 641} (2016) 1--68}, [\href{http://arxiv.org/abs/1603.04993}{{\tt
  1603.04993}}].

\bibitem{Coleman:1973jx}
S.~R. Coleman and E.~J. Weinberg, \emph{{Radiative Corrections as the Origin of
  Spontaneous Symmetry Breaking}},
  \href{http://dx.doi.org/10.1103/PhysRevD.7.1888}{\emph{Phys. Rev. D} {\bf 7}
  (1973) 1888--1910}.

\bibitem{BelleII:2023}
{S. Glazov, for the Belle II collaboration}, ``{Belle II Physics highlights}.''
  {talk at EPS-HEP August 24, 2023 Hamburg, Germany}.

\bibitem{Bause:2023mfe}
R.~Bause, H.~Gisbert and G.~Hiller, \emph{{Implications of an enhanced $B \to K
  \nu \bar \nu$ branching ratio}},  \href{http://arxiv.org/abs/2309.00075}{{\tt
  2309.00075}}.

\bibitem{Dreiner:2023cms}
H.~K. Dreiner, J.~Y. G\"unther and Z.~S. Wang, \emph{{The Decay $B\to
  K\nu\bar{\nu}$ at Belle II and a Massless Bino in R-parity-violating
  Supersymmetry}},  \href{http://arxiv.org/abs/2309.03727}{{\tt 2309.03727}}.

\bibitem{He:2023bnk}
X.-G. He, X.-D. Ma and G.~Valencia, \emph{{Revisiting models that enhance
  $B^+\to K^+ \nu\bar\nu$ in light of the new Belle II measurement}},
  \href{http://arxiv.org/abs/2309.12741}{{\tt 2309.12741}}.

\bibitem{Chen:2023wpb}
C.-H. Chen and C.-W. Chiang, \emph{{Flavor anomalies in leptoquark model with
  gauged $U(1)_{L_\mu-L_\tau}$}},  \href{http://arxiv.org/abs/2309.12904}{{\tt
  2309.12904}}.

\end{thebibliography}\endgroup

\end{document}